\documentclass{JHEP3}
\usepackage{latexsym}
\usepackage{graphics}

\def\comment #1{}
\def\cf {{\it cf. }}

\def\refer #1{{(\ref{#1})}}
\def\fullref #1{\ref{#1} (p.\pageref{#1})}

\def\ket #1{\left|{#1}\right\rangle}
\def\bracket #1#2{\left\langle{#1}|{#2}\right\rangle}

\def\of #1{\!\left({#1}\right)}

\def\Z {\theIntegers}
\def\N {\mathrm{I\!N}}
\def\R {\mathrm{I\!R}}
\def\C {\mathrm{\,I\!\!\!C}}

\def\grad {\nabla}
\def\gradOp {\hat\grad}

\def\set #1{\left\lbrace{#1}\right\rbrace}

\def\brackets #1{\left[{#1}\right]}
\def\braces #1{\left\lbrace{#1}\right\rbrace}

\def\commutator #1#2{\brackets{{#1},{#2}}}
\def\antiCommutator #1#2{\braces{{#1},{#2}}}
\def\superCommutator #1#2{\brackets{{#1},{#2}}_\involution}

\def\adjoint #1{{{{#1}^{\dag}}}}

\def\defas {:=}

\def\onbComp {\mathrm{e}}
\def\onbAnnihilator {\mbox{$\hat \onbComp$}}
\def\onbCreator {\mbox{$\adjoint{\onbAnnihilator}$}}
\def\coordAnnihilator {\mbox{$\hat \mathrm{c}$}}
\def\coordCreator {\adjoint{\coordAnnihilator}}

\def\manifold {\mathcal{M}}

\def\theIntegers {{\mathbf Z\!\!\!\mathbf Z}}
\def\extd {\mathbf{{d}}}

\def\coextd {\adjoint{\extd}}

\def\involution {\mathbf \iota}

\def\Dirac {\mathbf{D}}

\def\eigenspace #1#2 {\mathrm{eig}\of{#1,#2}}

\def\fatDelta {\mbox{\boldmath$\Delta$}}

\newlength{\skiplength}
\def\skiph #1{\settowidth{\skiplength}{#1}\hspace{\skiplength}}

\def\inner {\!\cdot\!}

\title{On deformations of 2d SCFTs}
\author{Urs Schreiber \\ Universit{\"a}t Duisburg-Essen \\ Essen, 45117, Germany\\
   E-mail: \email{Urs.Schreiber@uni-essen.de}}

\abstract{
Motivated by the representation of the super Virasoro constraints as 
generalized Dirac-K{\"a}hler constraints $(\extd \pm \coextd)\ket{\psi} = 0$  on loop space, 
examples of the most general continuous deformations $\extd \to e^{-\bf W}\,\extd\, e^{\bf W}$  
are considered which preserve the 
superconformal algebra
at the level of Poisson brackets. 
The deformations which induce the massless NS and NS-NS backgrounds are 
exhibited. 
Hints for a manifest realization of S-duality in terms of an algebra isomorphism are discussed. 

It is shown how the first order theory of 'canonical deformations'
is reproduced and how the deformation operator $\bf W$ encodes vertex operators and
gauge transformations. 
}

\begin{document}

\newpage
\tableofcontents

\newpage

\section{Introduction}

Supersymmetric field theories look like 
Dirac-K{\"a}hler systems when formulated in Schr{\"o}dinger representation.
This has been well studied in the special limits where only a finite number of
degrees of freedom are retained, such as the semi-classical quantization
of solitons in field theory (see e.g. \cite{HollowoodKingaby:2003}
for a brief introduction and further references). 
That this phenomenon is rooted in the general structure of supersymmetric field 
theory has been noted long ago in the second part of 
\cite{Witten:1982} (see also the second part of \cite{Witten:1985}). 
For 2 dimensional superconformal field theories describing
superstring worldsheets a way to exploit this fact for the construction of 
covariant target space Hamiltonians 
(applicable to the computation of curvature corrections of string spectra 
in nontrivial backgrounds) has been proposed in \cite{Schreiber:2003a}.
In the construction of these Hamiltonians a pivotal role is played 
by a new method for obtaining functional representations of
superconformal algebras (corresponding to non-trivial
target space backgrounds) by means of certain deformations of the 
superconformal algebra.

In \cite{Schreiber:2003a} the focus was on deformations which 
induce Kalb-Ramond backgrounds and
only the 0-mode of the superconformal algebra was considered explicitly (which
is sufficient for the construction of covariant target space Hamiltonians). Here
this deformation technique is developed in more detail for the full superconformal
algebra and for all massless bosonic string background fields. 
Other kinds of backgrounds can also be incorporated in principle and one goal 
of this paper is to demonstrate the versatility of the new deformation technique
for finding explicit functional realizations of the two-dimensional superconformal algebra.

The setting for our formalism is the
representation of the superconformal algebra on the exterior bundle
over loop space (the space of maps from the circle into target space) 
by means of $K$-deformed
exterior (co)derivatives $\extd_K$, $\coextd_K$, where
$K$ is the Killing vector field on loop space which induces loop reparameterizations.

The key idea is that 
the form of the superconformal algebra is preserved under the 
deformation\footnote{
  Throughout this paper we use the term ``deformation'' to mean the
  operation \refer{the fundamelental deformation idea} on the
  superconformal generators, the precise definition of which is given
  in \S\fullref{deformations}. These ``deformations'' are actually
  \emph{isomorphisms} of the superconformal algebra, but affect its
  representations in terms of operators on 
  the exterior bundle over loop space.
  In the literature one finds also other usages of the word ``deformation''
  in the context of superalgebras, 
  for instance for describing the map where the superbrackets 
  $\superCommutator{\cdot}{\cdot}$
  are transformed as
  \begin{eqnarray}
    \superCommutator{A}{B} \to \superCommutator{A}{B} + \sum\limits_{t=1}^\infty
    \omega_i\of{A,B}t^i
  \end{eqnarray}
  with $\omega_i\of{A,B}$ elements of the superalgebra and $t$ a real number 
  (see \cite{AzcarragaIzquierdoPiconVarela:2004}).
} 
\begin{eqnarray}
  \label{the fundamelental deformation idea}
  \extd_K &\to& e^{-\bf W}\,\extd_K \,e^{\bf W}
  \nonumber\\
  \coextd_K &\to& e^{{\bf W}^\dag}\,\coextd_K\, e^{-{\bf W}^\dag}
\end{eqnarray}
if $\bf W$ is an even graded operator that satisfies a certain consistency condition.

The canonical (functional) form of the superconformal generators for all 
massless NS and NS-NS backgrounds can neatly be expressed this way by deformation operators 
$\bf W$ that are bilinear in the fermions, as will be shown here. 
It turns out that there is one further bilinear in the fermions 
which induces a background that probably has to be interpreted as the RR 2-form as coupled to the
D-string.

It is straightforward to find further deformation operators and hence further backgrounds. 
While the normal ordering effects which affect the 
superconformal algebras and which would give rise to equations of motion for the
background fields are not investigated here,
there is still a consistency condition to be satisfied 
which constrains the admissible deformation operators. 

This approach for obtaining new superconformal algebras from existing ones by applying deformations 
is similar in spirit, but rather complementary, to the method of
'\emph{canonical deformations}' studied by Giannakis, Evans, Ovrut, Rama, Freericks, Halpern and  others
\cite{Giannakis:1999,OvrutRama:1992,EvansOvrut:1990,EvansOvrut:1989,FreericksHalpern:1988}. 
There, the superconformal generators $T$ and $G$ of one chirality are deformed to lowest order as
\begin{eqnarray}
  \label{canonical deformation in introduction}
  T\of{z} &\to& T\of{z} + \delta T\of{z}
  \nonumber\\
  G\of{z} &\to& G\of{z} + \delta G\of{z}
  \,.
\end{eqnarray}
Requiring the deformed generators to satisfy the desired algebra to first order shows 
that $\delta T$ and $\delta G$ must be bosonic and fermionic components 
of a weight 1 worldsheet superfield. 
(An adaption of this procedure to deformations of the BRST charge itself is discussed 
in \cite{Giannakis:2002}. Another related discussion of deformations of BRST operators
is given in \cite{Kato:1995}.)

The advantage of this method over the one discussed in the following is that it operates at the 
level of quantum SCFTs and has powerful CFT tools at its disposal, such as normal ordering and 
operator product expansion. The disadvantage is that it only applies perturbatively to 
first order in the background fields, and that these background fields always
appear with a certain gauge fixed.

On the other hand, the deformations discussed here which are induced by 
$\extd_K \to e^{-\bf W}\extd_K e^{\bf W} \sim e^{-\bf W}(iG + \bar G)e^{\bf W}$ 
preserve the superconformal algebra for arbitrarily large perturbations $\bf W$. 
The drawback is that normal ordering is non-trivially affected, too, and without 
further work the resulting superconformal algebra is only available  
on the level of (bosonic and fermionic) Poisson brackets. 

We show in \S\fullref{canonical deformations from d to dingens} that when
restricted to first order the deformations that we are considering reproduce the theory
of canonical deformations \refer{canonical deformation in introduction}.

Our deformation method is also technically different from but related to the 
\emph{marginal deformations} of conformal field theories 
(see \cite{FoersteRoggenkamp:2003} for a review and further references),
where one sends the correlation function $\langle A\rangle$ of some operator
$A$ to the deformed correlation function
\begin{eqnarray}
  \langle A\rangle^\lambda
  &\defas&
  \langle A \exp\of{\sum\limits_i \lambda_i \int {\cal O}_i\; {\rm dvol}}\rangle
  \,,
\end{eqnarray}
where ${\cal O}_i$ are fields of conformal weight $1$. This corresponds
to adding the integral over a field of unit weight to the action. How this relates to 
the algebraic deformations of the superconformal algebra considered here
is discussed in \S\fullref{review of first order canonical CFT deformations}.

The method discussed here generalizes the transformations 
studied in \cite{LizziSzabo:1998}, where strings are regarded from
the non-commutative geometry perspective. The main result of 
this approach (which goes back to \cite{FroehlichGawedzki:1993}
and \cite{EvansGiannakis:1996})
is that T-duality as well as mirror symmetry can nicely be encoded 
by means of automorphisms of the vertex operator algebra. In terms of
the above notation such automorphisms correspond to deformations
induced by \emph{anti-Hermitian} ${\bf W}^\dag = -{\bf W}$,
which induce pure gauge transformations on the algebra. 

The analysis given here generalizes the approach of \cite{LizziSzabo:1998}
in two ways: First, the use of Hermitian $\bf W$ in our formalism 
produces backgrounds which are not related by string dualities. Second, by 
calculating the functional form of the superconformal generators for
these backgrounds we can study the action of anti-Hermitian $\bf W$
on these more general generators and find the transformation of the
background fields under the associated target space duality.

In particular, we find a duality transformation which changes the sign of the
dilaton and interchanges $B$- and $C$-form fields. It would seem that
this must hence be related to S-duality. This question requires further analysis.

$\,$\\

The structure of this paper is as follows:

In \S\ref{loop space} some technical preliminaries necessary for
the following discussion are given.
The functional loop space notation is introduced in 
\S\ref{loop space definitions},
some basic facts about loop space geometry are discussed (\S\ref{exterior geometry on loop space}),
the exterior derivative and coderivative on that space are introduced 
(\S\ref{exterior algebra over loop space}),
and some remarks on isometries of loop space are given in \S\ref{isometries on loop space}.

This is then applied in \S\ref{loop space super-Virasoro generators} 
to the general analysis
of deformations of the superconformal generators. First of all,
the purely
gravitational target space background is shown to be associated to the 
ordinary $K$-deformed loop space exterior derivative (\S\ref{purely gravitational}).
\S\ref{deformations} 
then discusses how general continuous classical deformations
of the superconformal algebra are obtained. As a first application,
\S\ref{gravitational background by algebra isomorphism}
shows how this can be used to get the previously
discussed superconformal generators for purely gravitational backgrounds from
those of flat space by a deformation.

Guided by the form of this deformation the following sections
systematically list and analyze the deformations which are
associated with the Kalb-Ramond, dilaton, and gauge field backgrounds
(\S\ref{loop space and B-field background}, \S\ref{Dilaton background},
\S\ref{A-field background}).
It turns out (\S\ref{C-field background}) 
that one further 2-form background can be obtained
in a very similar fashion, which apparently has to be interpreted 
as the S-dual coupling of the D-string to the $C_2$ 2-form background.

After having understood how the NS-NS backgrounds arise in our formalism
we turn in \S\fullref{canonical deformations and vertex operators} to a comparison of the method 
presented here 
with the well-known 'canonical deformations', which are briefly reviewed in
\S\fullref{review of first order canonical CFT deformations}. 
In \S\fullref{canonical deformations from d to dingens} it is shown how these canonical deformations
are reproduced by means of the methods discussed here and how our deformation
operator $\bf W$ relates to the vertex operators of the respective background fields.

Next the inner relations between the various deformations found are
further analyzed in \S\ref{relations between susy algebras}. 
First of all \S\ref{dK exact deformations} demonstrates how
$\extd_K$-exact deformation operators yield target space gauge
transformations. Then, in \S\ref{T-duality} 
the well known realization of
T-duality as an algebra isomorphism is adapted to the present context,
and in \S\ref{T-duality for various background fields} 
the action of a target space duality obtained
from a certain modified algebra isomorphism on the various
background fields is studied. It turns out that there are
certain similarities to the action of loop space Hodge duality,
which is discussed in \S\ref{Hodge duality on loop space}.

Finally a summary and disucssion is given in 
\S\ref{summary and discussion}. 
The appendix lists some results from the canonical analysis of the
D-string action, which are needed in the main text.

\newpage
\section{Loop space}
\label{loop space}

In this section the technical setup is briefly established. The  
0-mode $\extd_K$ of the sum of the left- and the rightmoving 
supercurrents is represented as the $K$-deformed exterior derivative on loop space. 
Weak nilpotency of this $K$-deformed operator (namely nilpotency up to reparameterizations) 
is the essential property which implies that the modes of $\extd_K$ and its adjoint generate 
a superconformal algebra. In this sense the loop space perspective on superstrings highlights 
a special aspect of the super Virasoro constraint algebra which turns out to be pivotal 
for the construction of classical deformations of that algebra. 

The kinematical configuration space of the closed \emph{bosonic} string is loop space
${\cal L}\manifold$, the
space of parameterized loops in target space $\manifold$. 
As discussed in \S 2.1 of \cite{Schreiber:2003a}
the kinematical configuration space of the closed \emph{superstring} is therefore the superspace
over ${\cal L}\manifold$, which can be identified with the 1-form bundle
$\Omega^1\of{{\cal L}\manifold}$. Superstring states in Schr{\"o}dinger representation
are super-functionals on $\Omega^1\of{{\cal L}\manifold}$ and hence section of the form bundle
$\Omega\of{{\cal L}\manifold}$ over loop space. 

The main
technical consequence of the infinite dimensionality are the well known divergencies
of certain objects, such as the Ricci-Tensor and the Laplace-Beltrami operator, which
inhibit the naive implementation of quantum mechanics on ${\cal L}\manifold$. But of course
these are just the well known infinities that arise, when working in the Heisenberg (CFT)
instead of in the Schr{\"o}dinger picture, from operator ordering effects, and which
should be removed by imposing normal ordering. Since the choice of Schr{\"o}dinger or Heisenberg
picture is just one of language, the same normal ordering (now expressed in terms of 
functional operators instead of Fock space operators) takes care of infinities in
loop space. We will therefore not have much more to say about this issue here. The
main result of this section are various (deformed) 
representations of the super-Virasoro algebra on loop space
(corresponding to different spacetime backgrounds), and will be
derived in their classical
(Poisson-bracket) form without considering normal ordering effects.

A mathematical discussion of aspects of
loop space can for instance be found in \cite{Wurzbacher:1995,Wurzbacher:2001}.
A rigorous treatment of some of the objects discussed below is also given in
\cite{Arai:1993}.

\subsection{Definitions}
\label{loop space definitions}

Let $\left(\manifold,g\right)$ be a pseudo-Riemannian manifold, the target space,
 with metric $g$,  and let
${\cal L}\manifold$ be its loop space consisting of smooth maps of the \emph{parameterized}
circle with parameter $\sigma \sim \sigma + 2\pi$ into $\manifold$:
\begin{eqnarray}
  {\cal L}\manifold &\defas& C^\infty\of{S^1,\manifold} 
  \,.
\end{eqnarray}
The tangent space $T_X{\cal L}\manifold$ of ${\cal L}\manifold$ 
at a loop $X : S^1 \to \manifold$ is  the space of vector fields along that loop.
The metric on $\manifold$ induces a metric on $T_X{\cal L}\manifold$:
Let $g\of{p} = g_{\mu\nu}\of{p}dx^\mu \otimes dx^\nu$ 
be the metric tensor on $\manifold$.
Then we choose for the metric on ${\cal L}\manifold$ at a point $X$ the mapping
\begin{eqnarray}
  \label{metric on loop space}
  T_X{\cal L}\manifold \times T_X{\cal L}\manifold
 &\to&
  \R
  \nonumber\\
  (U,V) 
  &\mapsto&
  U\inner V
  \;=\;
  \int\limits_0^{2\pi} d\sigma\; g\of{X\of{\sigma}}\of{U\of{\sigma}, V\of{\sigma}}
  \nonumber\\
  &&
  \skiph{$U\inner V\;$}
  \;=\;
  \int\limits_0^{2\pi} d\sigma\; g_{\mu\nu}\of{X\of{\sigma}}U^\mu\of{\sigma}V^\nu\of{\sigma}
  \,.
\end{eqnarray}
For the intended applications $T{\cal L}\manifold$ is actually too small, since there will be need to deal with
distributional vector fields on loop space. Therefore one really considers 
$\bar T{\cal L}\manifold$, the completion of $T{\cal L}\manifold$ at each point
$X$ with respect to the norm induced by the inner product \refer{metric on loop space}.)
For brevity, whenever we refer to ``loop space'' in the following, we mean ${\cal L}\manifold$
equipped with the metric \refer{metric on loop space}. Also, the explicit integration region 
$\sigma \in (0,2\pi)$ will be implicit in the following.

To abbreviate the notation, let us introduce formal multi-indices $(\mu,\sigma)$ and
write equivalently 
\begin{eqnarray}
  \label{loop space multi-indices}
  U^\mu\of{\sigma}  
  &\defas&
  U^{(\mu,\sigma)}
\end{eqnarray}
for a vector $U\in T_X{\cal L}\manifold$, and similarly for higher-rank tensors on
loop space.

Extending the usual index notation to the infinite-dimensional setting in the 
obvious way, we also write:
\begin{eqnarray}
  \int U^\mu\of{\sigma} V_\mu\of{\sigma}
  &\defas&
  U^{(\mu,\sigma)}V_{(\mu,\sigma)}
  \,.
\end{eqnarray}
For this to make sense we need to know how to ``shift'' the continuous index
$\sigma$. Because of
\begin{eqnarray}
  \int d\sigma \; g_{\mu\nu}\of{X\of{\sigma}}U^\mu\of{\sigma}V^\nu\of{\sigma}
   &=&
  \int d\sigma\;d\sigma^\prime \;
    \delta\of{\sigma,\sigma^\prime}
   g_{\mu\nu}\of{X\of{\sigma}}U^\mu\of{\sigma}V^\nu\of{\sigma^\prime}
  \nonumber
\end{eqnarray}
it makes sense to write the metric tensor on loop space as
\begin{eqnarray}
  \label{concise metric on loop space}
  G_{(\mu,\sigma)(\nu,\sigma^\prime)}\of{X}
  &\defas&
  g_{\mu\nu}\of{X\of{\sigma}}\delta\of{\sigma,\sigma^\prime}
  \,.
\end{eqnarray}
Therefore 
\begin{eqnarray}
  \langle U, V\rangle
  &=&
  U^{(\mu,\sigma)}
  G_{(\mu,\sigma)(\nu,\sigma^\prime)}
  V^{(\nu,\sigma^\prime)}
\end{eqnarray}
and
\begin{eqnarray}
  V_{(\mu,\sigma)}
  &=&
  G_{(\mu,\sigma)(\nu,\sigma^\prime)}
  V^{(\nu,\sigma^\prime)}
  \nonumber\\
  &=&
  V_\mu\of{\sigma}
  \,.
\end{eqnarray}
Consequently, it is natural to write
\begin{eqnarray}
  \delta\of{\sigma,\sigma^\prime}
  &\defas&
  \delta^{\sigma^\prime}_\sigma 
  \;=\; \delta^\sigma_{\sigma^\prime} 
  \;=\;\delta_{\sigma,\sigma^\prime} \;=\;\delta^{\sigma,\sigma^\prime}
  \,.
\end{eqnarray}
A (holonomic) basis for $T_X{\cal L }\manifold$ may now be written as
\begin{eqnarray}
  \partial_{(\mu,\sigma)}
  &\defas&
  \frac{\delta}{\delta X^\mu\of{\sigma}}
  \,,
\end{eqnarray}
where the expression on the right denotes the functional derivative, so that
\begin{eqnarray}
  \partial_{(\mu,\sigma)} X^{(\nu,\sigma^\prime)}
  &=&
  \delta^{(\nu,\sigma^\prime)}_{(\mu,\sigma)}
  \nonumber\\
  &=&
  \delta^\nu_\mu\,\delta\of{\sigma,\sigma^\prime}
  \,.
\end{eqnarray}

By analogy, many concepts known from finite dimensional geometry carry over
to the infinite dimensional case of loop spaces. Problems arise when 
traces over the continuous ``index'' $\sigma$ are taken, like for contractions
of the Riemann tensor, which leads to undefined diverging expressions. 
It is expected that these are taken care of by the usual normal-ordering of quantum field theory.

\subsection{Differential geometry on loop space}
\label{exterior geometry on loop space}

With the metric \refer{concise metric on loop space} on loop space in hand
\begin{eqnarray}
  \label{metric on loop space again}
  G_{(\mu,\sigma)(\nu,\sigma^\prime)}\of{X}
  &=&
  g_{\mu\nu}\of{X\of{\sigma}}\delta_{\sigma,\sigma^\prime}
\end{eqnarray}
the usual objects of differential geometry can be derived for loop space.
Simple calculations yield the Levi-Civita connection as well as the
Riemann curvature, which will be frequently needed later on. The exterior
algebra over loop space is introduced and the exterior derivative and its
adjoint,
which play the central role in the construction
of the super-Virasoro algebra in \S\fullref{purely gravitational},
are constructed in terms of operators on the exterior bundle.
Furthermore isometries on loop space are considered, both 
the one coming from reparameterization of loops as well as those induced from
target space. The former leads to the reparameterization constraint on strings,
while the latter is crucial for the Hamiltonian evolution on loop space
\cite{Schreiber:2003a}.

\subsubsection{Basic geometric data.}
The inverse metric is obviously
\begin{eqnarray}
  G^{(\mu,\sigma)(\nu,\sigma^\prime)}\of{X}
  &=&
  g^{\mu\nu}\of{X\of{\sigma}} \delta\of{\sigma,\sigma^\prime}
  \,.
\end{eqnarray}
A vielbein field ${\bf e}^a = e^a{}_\mu \extd x^\mu$ on $\manifold$ gives rise
to a vielbein field ${\bf E}^{(a,\sigma)}$ on loop space:
\begin{eqnarray}
  \label{vielbein on loop space}
  E^{(a,\sigma)}{}_{(\mu,\sigma^\prime)}\of{X}
  &\defas&
  e^a{}_\mu\of{X\of{\sigma}}\delta^\sigma{}_{\sigma^\prime}
\end{eqnarray}
which satisfies
\begin{eqnarray}
  E^{(a,\sigma)}{}_{(\mu,\sigma^{\prime\prime})}
  E^{(b,\sigma)(\mu,\sigma^{\prime\prime})}
  &=&
  \eta^{ab}\delta^{\sigma,\sigma^\prime}
  \nonumber\\
  &\defas&
  \eta^{(a,\sigma)(b,\sigma^\prime)}
\end{eqnarray}
For the Levi-Civita connection one finds:
\begin{eqnarray}
  &&\Gamma_{(\mu\sigma)(\alpha \sigma^\prime)(\beta\sigma^{\prime\prime})}\of{X}
  \nonumber\\
  &=&
  \frac{1}{2}
  \left(
    \frac{\delta}{\delta X^\mu\of{\sigma}} G_{(\alpha,\sigma^\prime)(\beta,\sigma^{\prime\prime})}\of{X}
    +
    \frac{\delta}{\delta X^\beta\of{\sigma^{\prime\prime}}} G_{(\mu,\sigma)(\alpha,\sigma^\prime)}\of{X}
    -
    \frac{\delta}{\delta X^\alpha\of{\sigma^{\prime}}} G_{(\beta,\sigma^{\prime\prime})(\mu,\sigma)}\of{X}
  \right)
  \nonumber\\
  &=&
  \frac{1}{2}\left(
  (\partial_\mu G_{\alpha\beta})\of{X\of{\sigma^\prime}}
  \delta\of{\sigma,\sigma^\prime}\delta\of{\sigma^\prime,\sigma^{\prime\prime}}
  +
  (\partial_\beta G_{\mu\alpha})\of{X\of{\sigma}}
  \delta\of{\sigma^{\prime\prime},\sigma}\delta\of{\sigma,\sigma^{\prime}}  
  \right)
  \nonumber\\
  &&
  -
  \frac{1}{2}
  (\partial_\alpha G_{\beta\mu})\of{X\of{\sigma^{\prime\prime}}}
  \delta\of{\sigma^{\prime},\sigma^{\prime\prime}}\delta\of{\sigma^\prime,\sigma}  
  \nonumber\\
  &=&
  \Gamma_{\mu\alpha\beta}\of{X\of{\sigma}}\;
     \delta\of{\sigma,\sigma^\prime}\delta\of{\sigma^\prime,\sigma^{\prime\prime}}
  \,,
\end{eqnarray}
and hence
\begin{eqnarray}
  \label{loop space LC connection}
  \Gamma_{(\mu,\sigma)}{}^{(\alpha,\sigma^\prime)}{}_{(\beta,\sigma^{\prime\prime})}
  \of{X}
  &=&
  \Gamma_{\mu}{}^\alpha{}_\beta\of{X\of{\sigma}}\;
     \delta\of{\sigma,\sigma^\prime}\delta\of{\sigma^\prime,\sigma^{\prime\prime}}  
 \,.
\end{eqnarray}
The respective connection in an orthonormal basis is
\begin{eqnarray}
  \omega_{(\mu,\sigma)}{}^{(a\sigma^\prime)}{}_{(b,\sigma^{\prime\prime})}\of{X}
  &=&
  E^{(a,\sigma^\prime)}{}_{(\alpha,\rho)}\of{X}
  \left(
    \delta^{(\alpha,\rho)}_{(\beta,\rho^\prime)}
    \partial_{(\mu,\sigma)}
    +
    \Gamma_{(\mu,\sigma)}{}^{(\alpha,\rho)}{}_{(\beta,\rho^{\prime})}\of{X}
  \right)
  E^{(\beta,\rho^{\prime})}{}_{(b,\sigma^\prime)}\of{X}
  \nonumber\\
  &=&
  \omega_\mu{}^a{}_b\of{X\of{\sigma}}
  \delta\of{\sigma,\sigma^\prime}\delta\of{\sigma^\prime,\sigma^{\prime\prime}}
  \,.
\end{eqnarray}
From \refer{loop space LC connection} the Riemann tensor on loop space is obtained as
\begin{eqnarray}
  &&R_{(\mu,\sigma_1)(\nu,\sigma_2)}{}^{(\alpha,\sigma_3)}{}_{(\beta,\sigma_4)}\of{X}
  \nonumber\\
  &=&
  2
  \frac{\delta}{\delta X^{[(\mu,\sigma_1)}}
  \Gamma_{(\nu,\sigma_2)]}{}^{(\alpha,\sigma_3)}{}_{(\beta,\sigma_4)}
  +
  2
  \Gamma_{[(\mu,\sigma_1)}{}^{(\alpha,\sigma_3)}{}_{|(X,\sigma_5)|}
  \Gamma_{(\nu,\sigma_2)]}{}^{(X,\sigma_5)}{}_{(\beta,\sigma_4)}
  \nonumber\\
  &=&
  R_{\mu\nu}{}^\alpha{}_\beta\of{X\of{\sigma_1}}\;
  \delta\of{\sigma_1,\sigma_2}
  \delta\of{\sigma_2,\sigma_3}
  \delta\of{\sigma_3,\sigma_4}
  \,.
\end{eqnarray}
The Ricci tensor is formally
\begin{eqnarray}
  R_{(\mu,\sigma)(\nu,\sigma^\prime)}\of{X}
  &=&
  R_{(\kappa,\sigma^{\prime\prime})(\mu,\sigma)}{}^{(\kappa,\sigma^{\prime\prime})}{}_{(\nu,\sigma^\prime)}\of{X}
  \nonumber\\
  &=&
  R_{\mu\nu}\of{X\of{\sigma}}\delta\of{\sigma,\sigma^\prime} \;\delta\of{\sigma^{\prime\prime},\sigma^{\prime\prime}}
  \,,
\end{eqnarray}
which needs to be regularized. Similarly the curvature scalar is formally
\begin{eqnarray}
  R\of{X}
  &=&
  R_{(\mu,\sigma)}{}^{(\mu,\sigma)}\of{X}
  \nonumber\\
  &=&
  R\of{X\of{\sigma}} \delta^\sigma_\sigma \delta\of{\sigma^{\prime\prime},\sigma^{\prime\prime}}
  \,.
\end{eqnarray}

\subsubsection{Exterior and Clifford algebra over loop space.}
\label{exterior algebra over loop space}

The anticommuting fields ${\cal E}^{\dag (\mu,\sigma)}$,
${\cal E}_{(\mu,\sigma)}$, satisfying the CAR
\begin{eqnarray}
 \antiCommutator
   {{\cal E}^{\dag(\mu,\sigma)}}
   {{\cal E}^{\dag (\nu,\sigma^\prime)}}
  &=& 0
 \nonumber\\
 \antiCommutator
   {{\cal E}_{(\mu,\sigma)}}
   {{\cal E}_{(\nu,\sigma^\prime)}}
  &=& 0
 \nonumber\\
 \antiCommutator
   {{\cal E}_{(\mu,\sigma)}}
   {{\cal E}^{\dag (\nu,\sigma^\prime)}}
  &=&
  \delta^{(\mu,\sigma)}_{(\nu,\sigma^\prime)}
  \,,
\end{eqnarray}
are assumed to exist over loop space, in analogy with
the creators and annihilators $\coordCreator^\mu$, $\coordAnnihilator_\mu$ 
on the exterior bundle in finite dimensions as described in  
appendix A of \cite{Schreiber:2003a}.
(For a mathematically rigorous treatment of the continuous CAR
compare \cite{Wurzbacher:2001} and references given there.)
From them the Clifford fields
\begin{eqnarray}
  \label{loop space cliffords}
  \Gamma_\pm^{(\mu,\sigma)}
  &\defas&
  {\cal E}^{\dag (\mu,\sigma)}
  \pm{\cal E}^{(\mu,\sigma)}
\end{eqnarray}
are obtained, which satisfy
\begin{eqnarray}
  \label{modes by commuting with number operator}
  \antiCommutator{\Gamma_\pm^{(\mu,\sigma)}}{\Gamma_\pm^{(\nu,\sigma^\prime)}}
  &=&
 \pm 2 G^{(\mu,\sigma)(\nu,\sigma^\prime)}
  \nonumber\\
  \antiCommutator{\Gamma_\pm^{(\mu,\sigma)}}{\Gamma_\mp^{(\nu,\sigma^\prime)}}
  &=&
  0
  \,.
\end{eqnarray}
Since the $\Gamma_\pm$ will be related to spinor fields on the string's worldsheet,
we alternatively use spinor indices $A, B,\ldots \in \set{1,2} \simeq \set{+,-}$ and write
\begin{eqnarray}
  \antiCommutator{
   \Gamma_A^{(\mu,\sigma)}}{\Gamma_B^{(\nu,\sigma^\prime)}}
 &=&
  2 s_A \delta_{AB} G^{(\mu,\sigma)(\nu,\sigma^\prime)}
  \,.
\end{eqnarray}
Here $s_A$ is defined by
\begin{eqnarray}
 s_+ \;=\; +1\,,\hspace{.2cm} s_-\;=\;-1
  \,.
\end{eqnarray}
The above operators will frequently be needed with respect to some orthonormal frame
$E^{(a,\sigma)}$:
\begin{eqnarray}
  \Gamma_A^{(a,\sigma)}
  &\defas&
  E^{(a,\sigma)}{}_{(\mu,\sigma^\prime)}\Gamma_A^{(\mu,\sigma^\prime)}
  \,.
\end{eqnarray}

Just like in the finite dimensional case, the following derivative operators
can now be defined:

The covariant derivative operator (\cf A.2 in \cite{Schreiber:2003a}) on 
the exterior bundle over loop space is
\begin{eqnarray}
  \gradOp_{(\mu,\sigma)}
  &=&
  \partial^c_{(\mu,\sigma)}
  -
  \Gamma_{(\mu,\sigma)}{}^{(\alpha,\sigma^\prime)}{}_{(\beta,\sigma^{\prime\prime})}
  {\cal E}^{\dag(\beta,\sigma^{\prime\prime})}
  {\cal E}_{(\alpha,\sigma^\prime)}
  \nonumber\\
  &=&
  \partial^c_{(\mu,\sigma)}
  -
  \int d\sigma^\prime\;d\sigma^{\prime\prime}
  \Gamma_{\mu}{}^\alpha{}_\beta\of{X\of{\sigma}}
  \delta\of{\sigma,\sigma^\prime}
  \delta\of{\sigma^\prime,\sigma^{\prime\prime}}
  {\cal E}^{\dag \beta}\of{\sigma^{\prime\prime}}
  {\cal E}_{\alpha}\of{\sigma^\prime}
  \nonumber\\
  &=&
  \partial^c_{(\mu,\sigma)}
  -
  \Gamma_{\mu}{}^\alpha{}_\beta\of{X\of{\sigma}}
  {\cal E}^{\dag \beta}\of{\sigma}
  {\cal E}_{\alpha}\of{\sigma}
\end{eqnarray}
or alternatively
\begin{eqnarray}
  \label{loop space grad op}
  \gradOp_{(\mu,\sigma)}
  &=&
  \partial_{(\mu,\sigma)}
  -
  \omega_{\mu}{}^a{}_b\of{X\of{\sigma}}
  {\cal E}^{\dag b}\of{\sigma}
  {\cal E}_{a}\of{\sigma}  
  \,.
\end{eqnarray}
One should note well the difference between the functional derivative 
$\partial^c_{(\mu,\sigma)}$ which commutes with the coordinate frame
forms ($[{\partial^c_{(\mu,\sigma)}}{{\cal E}^{\dag \nu}}] = 0$)
and the functional derivative $\partial_{(\mu,\sigma)}$ which instead
commutes with the ONB frame forms
($[{\partial_{(\mu,\sigma)}}{{\cal E}^{\dag a}}] = 0$). See
(A.29) of \cite{Schreiber:2003a} for more details.

In terms of these operators the exterior derivative and coderivative on loop space read, 
respectively (A.39)
\begin{eqnarray}
  \label{extds on loop space}
  \extd
  &=&
  {\cal E}^{\dag(\mu,\sigma)}\partial^c_{(\mu,\sigma)}
  \nonumber\\
  &=&
  {\cal E}^{\dag(\mu,\sigma)}\gradOp_{(\mu,\sigma)}
  \nonumber\\
  \coextd
  &=&
  -{\cal E}^{(\mu,\sigma)}\gradOp_{(\mu,\sigma)}
  \,.
\end{eqnarray}

We will furthermore need the form number operator
\begin{eqnarray}
  {\cal N}
  &=&
  {\cal E}^{\dag (\mu,\sigma)}{\cal E}_{(\mu,\sigma)}
\end{eqnarray}
as well as its \emph{modes}: Let $\xi : S^1 \to \C$ be a smooth function then
\begin{eqnarray}
  \label{mode of form number operator}
  {\cal N}_\xi &\defas&
  \int d\sigma\;
  \xi\of{\sigma} {\cal E}^{\dag \mu}\of{\sigma}{\cal E}_{\mu}\of{\sigma}
\end{eqnarray}
is the $\xi$-mode of the form number operator. Commuting it with the
exterior derivative yields the modes of that operator:
\begin{eqnarray}
  \extd_\xi
  &\defas&
  \commutator{{\cal N}_\xi}{\extd}
  \nonumber\\
  &=&
  \int d\sigma\;
  \xi\of{\sigma}
  \;
  {\cal E}^{\dag \mu}\of{\sigma}\gradOp_\mu\of{\sigma}
  \nonumber\\
  \coextd_\xi &\defas&
  -
  \commutator{{\cal N}_\xi}{\coextd}
  \nonumber\\
  &=&
  -
  \int d\sigma\;
  \xi\of{\sigma}
  \;
  {\cal E}^{\mu}\of{\sigma}\gradOp_\mu\of{\sigma}
  \,.
\end{eqnarray}
These modes will play a crucial role in \S\fullref{loop space super-Virasoro generators}.

\subsubsection{Isometries.}
\label{isometries on loop space}

Regardless of the symmetries of the metric $g$ on $\manifold$, loop space
$\left({\cal L}\manifold,G\right)$ has an isometry generated by the
reparameterization flow vector field $K$, which is defined by:\footnote{
Here and in the following a prime indicates the derivative with respect to
the loop parameter $\sigma$: $X^\prime\of{\sigma} = \partial_\sigma X\of{\sigma}$.
}
\begin{eqnarray}
  \label{reparameterization Killing field on loop space}
  K^{(\mu,\sigma)}\of{X}
  &=&
  T
  \,
  X^{\prime\mu}\of{\sigma}
  \,.
\end{eqnarray}
(Here $T$ is just a constant which we include for later convenience.)
The flow generated by this
vector field obviously rotates the loops around. Since the metric \refer{metric on loop space again}
is ``diagonal'' in the parameter $\sigma$, this leaves the geometry of loop space invariant, and
the vector field $K$ satisfies Killing's equation
\begin{eqnarray}
  G_{(\nu,\sigma^\prime)(X,\sigma^{\prime\prime})}
  \nabla_{(\mu,\sigma)} K^{(X,\sigma^{\prime\prime})}
  +
  G_{(\mu,\sigma)(X,\sigma^{\prime\prime})}
  \nabla_{(\nu,\sigma^\prime)} K^{(X,\sigma^{\prime\prime})}
  &=&
  0\,,
\end{eqnarray}
as is readily checked.

The Lie-derivative along $K$ is (see section A.4 of \cite{Schreiber:2003a})
\begin{eqnarray}
  {\cal L}_K
  &=&
  \antiCommutator{{\cal E}^{\dag(\mu,\sigma)}\partial^c_{(\mu,\sigma)}}
  {{\cal E}_{(\nu,\sigma^\prime)}X^{\prime (\nu,\sigma^\prime)}}
  \nonumber\\
  &=&
  X^{\prime(\mu,\sigma)}
  \partial^c_{(\mu,\sigma)}
  +
  {\cal E}^{\dag(\mu,\sigma)}{\cal E}_{(\nu,\sigma^\prime)}
  \delta^\prime_{\sigma^\prime,\sigma}
  \nonumber\\
  &=&
  X^{\prime(\mu,\sigma)}
  \partial^c_{(\mu,\sigma)}
  +
  {\cal E}^{\dag\prime(\mu,\sigma)}{\cal E}_{(\mu,\sigma)}
  \,.
\end{eqnarray}
This operator will be seen to be an essential ingredient of the 
super-Virasoro algebra in \S\fullref{loop space super-Virasoro generators}.

Apart from the generic isometry \refer{reparameterization Killing field on loop space},
every symmetry of the target space manifold $\manifold$ gives rise to a family
of symmetries on ${\cal L}\manifold$:
Let $v$ be any Killing vector on target space,
\begin{eqnarray}
  \nabla_{(\mu}v_{\nu)} &=& 0
  \,,
\end{eqnarray}
then every vector $V$ on loop space of the form
\begin{eqnarray}
  \label{loop space Killing from target space Killing}
  V_\xi\of{X} 
  &=& V_\xi^{(\mu,\sigma)}\of{X}\partial_{(\mu,\sigma)}
  \;\defas\;
  v^\mu\of{X\of{\sigma}} \xi^\sigma \partial_{(\mu,\sigma)}
  \,,
\end{eqnarray}
where $\xi^\sigma = \xi\of{\sigma}$ is some differentiable function $S^1\to \C$,
is a Killing vector on loop space.
For the commutators one finds
\begin{eqnarray}
  \label{commutator of reparameterization with induced Killing vectors}
  \commutator{V_{\xi_1}}{V_{\xi_2}} &=& 0
  \nonumber\\
  \commutator{V_\xi}{K} &=& V_{\xi^\prime}
  \,.
\end{eqnarray}
The reparameterization Killing vector $K$ will be used to deform
the exterior derivative on loop space as discussed in \S 2.1.1 of 
\cite{Schreiber:2003a}, and a target space induced Killing vector $V_\xi$ will
serve as a generator of parameter evolution on loop space along the lines
of \S 2.2 of \cite{Schreiber:2003a}. There it was found in equation (88) that the
condition
\begin{eqnarray}
  \commutator{K}{V_\xi} &=& 0
\end{eqnarray}
needs to be satisfied for this to work. Due to 
\refer{commutator of reparameterization with induced Killing vectors}
this means that one needs to choose $\xi = {\rm const}$, i.e. use the
integral lines of $V_{\xi =1 }$ as the ``time''-parameter on loop space.
This is only natural: It means that every point on the loop is evolved
equally along the Killing vector field $v$ on target space.

\section{Superconformal generators for various backgrounds}
\label{loop space super-Virasoro generators}

We now use the loop space technology to 
show that the loop space exterior derivative deformed by the
reparameterization Killing vector $K$ gives rise to the superconformal
algebra which describes string propagation in purely gravitational
backgrounds. General deformations of this algebra are introduced and
applying these we find representations
of the superconformal algebra that correspond to all the massless
NS and NS-NS background fields.

(Parts of this construction were already indicated in \cite{Schreiber:2003a},
but there only the 0-modes of the generators and 
only a subset of massless bosonic background fields was considered, without 
spelling out the full nature of the necessary constructions on loop space.)

\subsection{Purely gravitational background}
\label{purely gravitational}

In this subsection it is described how to obtain a representation of the classical super-Virasoro algebra
on loop space. For a trivial background the construction itself is relatively trivial and,
possibly in different notation, well known. The point that shall be emphasized here
is that the identification of super-Virasoro generators with modes of the 
deformed exterior(co-)derivative on loop space allows a convenient treatment of curved
backgrounds as well as more general non-trivial background fields.

As was discussed in \cite{Schreiber:2003a}, \S 2.1.1 
(which is based on \cite{Witten:1982,Witten:1985}), 
one may obtain from the exterior derivative and
its adjoint on a manifold the generators of a global $D=2$, $N=1$ superalgebra by
deforming with a Killing vector. The generic Killing vector field on loop space is
the reparameterization generator \refer{reparameterization Killing field on loop space}.
Using this to deform the exterior derivative and its adjoint as in 
equation (19) of \cite{Schreiber:2003a} yields the operators
\begin{eqnarray}
  \label{k-defomred extds on loop space}
  \extd_{K}
  &\defas&
  \extd + i{\cal E}_{(\mu,\sigma)}X^{\prime(\mu,\sigma)}
  \nonumber\\
  \coextd_{K}
  &\defas&
  \coextd - i{\cal E}^\dag_{(\mu,\sigma)}X^{\prime(\mu,\sigma)}  
  \,,
\end{eqnarray}
(where for convenience we set $T=1$ for the moment)
which generate a \emph{global} superalgebra. Before having a closer look at this algebra
let us first enlarge it to a local superalgebra by considering the \emph{modes}
defined by
\begin{eqnarray}
  \label{super-virasoro generators}
  \extd_{K,\xi}
  &\defas&
  \commutator{{\cal N}_\xi}{\extd^\ast_{K}}
  \nonumber\\
  \coextd_{K,\xi^\ast}
  &\defas&
  -\commutator{{\cal N}_\xi}{\coextd^\ast_{K}}  
  \,,
\end{eqnarray}
where $\cdot^\ast$ is the complex adjoint and ${\cal N}_\xi$ is the $\xi$-mode of the
form number operator discussed in \refer{mode of form number operator}.
They explicitly read
\begin{eqnarray}
  \label{modes of deformed extd coexted on loop space}
  \extd_{K,\xi}
  &=&
  \int d\sigma\;
  \xi\of{\sigma}
  \left(
    {\cal E}^{\dag\mu}\of{\sigma}
    \partial^c_\mu\of{\sigma}
    +
    i{\cal E}_\mu\of{\sigma}
    X^{\prime \mu}\of{\sigma}
  \right)
  \nonumber\\
  \coextd_{K,\xi}
  &=&
  -
  \int d\sigma\;
  \xi\of{\sigma}
  \left(
    {\cal E}^{\mu}\of{\sigma}
    \nabla_\mu\of{\sigma}
    +
    i{\cal E}^\dag_\mu\of{\sigma}
    X^{\prime \mu}\of{\sigma}
  \right)
  \,.  
\end{eqnarray}
Making use of the fact that $\extd_{K,\xi}$ is actually independent of the background metric,
it is easy to establish the algebra of these operators. We do this for the ``classical'' fields,
ignoring normal ordering effects and the anomaly:

The anticommutator  of the operators \refer{super-virasoro generators}
with themselves
defines the $\xi$-mode ${\cal L}_{K,\xi}$ of the Lie-derivative ${\cal L}_{K}$ along $K$:
\begin{eqnarray}
  \label{anticom of extds}
  \antiCommutator{\extd_{K,\xi_1}}{\extd_{K,\xi_2}}  
  &=&
  2i{\cal L}_{K,\xi_1\xi_2}
  \,,
\end{eqnarray}
where
\begin{eqnarray}
  \label{modes of the reparameterization Lie derivative}
  {\cal L}_{\xi}
  &=&
  \int d\sigma\;
  \left(
  \xi\of{\sigma}
    X^{\prime\mu}\of{\sigma}
   \partial_\mu^c\of{\sigma}
   +
   \sqrt{\xi}
   \left(
     \sqrt{\xi}
     {\cal E}^{\dag\mu}
   \right)^\prime
  \of{\sigma}
   {\cal E}_\mu\of{\sigma}
  \right)
  \,.
\end{eqnarray}
We say that a field $A\of{\sigma}$ has \emph{reparameterization weight}
$w$ if
\begin{eqnarray}
  \label{transformation under reparameterization}
  \superCommutator{{\cal L}_{\xi}}{A\of{\sigma}}
  &=&
  \left(
  \xi A^\prime + w \xi^\prime A
  \right)\of{\sigma}
  \nonumber\\
  \superCommutator{{\cal L}_{\xi_1}}{A_{\xi_2}}
  &=&
  A_{(w-1)\xi_1^\prime\xi_2 - \xi_1\xi_2^\prime}
  \,,
\end{eqnarray}
where $A_\xi \defas \int d\sigma\; \xi A$.
For the basic fields we find
\begin{eqnarray}
  \label{reparameterization weight of basic fields on loop space}
  w\of{X^\mu} &=& 0
  \nonumber\\
  w\of{X^{\prime\mu}} &=& 1
  \nonumber\\
  w\of{\partial^c_\mu} &=& 1
  \nonumber\\
  w\of{\Gamma_\pm^\mu} &=& 1/2
  \,.
\end{eqnarray}
Because of  $w\of{AB} = w\of{A} + w\of{B}$ it follows that
$\extd_{K,\xi}$ and $\coextd_{K,\xi}$ are modes of integrals over densities of
reparameterization weight $w = 3/2$. This implies in particular that
\begin{eqnarray}
  \label{comm of L extd}
  \commutator
   {{\cal L}_{\xi_1}}
   {\extd_{K,{\xi_2}}}
   &=&
  \extd_{K,(\frac{1}{2}\xi_1^\prime\xi_2 - \xi_1\xi_2^\prime)}
  \\
  \commutator{{\cal L}_{K,\xi_1}}{{\cal L}_{K,\xi_2}}
  &=&
  {\cal L}_{K,(\xi_1^\prime\xi_2 - \xi_1\xi_2^\prime)}
  \,.
\end{eqnarray}

By taking the adjoint of \refer{anticom of extds} and \refer{comm of L extd}
(or by doing the calculation explicitly),
analogous relations are found for $\coextd_{K,\xi}$:
\begin{eqnarray}
  \label{vir algebra for coextd}
  \antiCommutator{\coextd_{K,\xi_1}}{\coextd_{K,\xi_2}}
  &=&
  2i {\cal L}_{K,\xi_1\xi_2}
  \nonumber\\
  \label{comm L coextd}
  \commutator{
    {\cal L}_{K,\xi_1}
  }{\coextd_{K,\xi_2}}
  &=&
  \coextd_{K,(\frac{1}{2}\xi_1^\prime\xi_2-\xi_1\xi_2^\prime)}
  \,.
\end{eqnarray}
Equations \refer{anticom of extds}, 
\refer{comm of L extd}, and \refer{vir algebra for coextd} 
give us part of the sought-after algebra.
A very simple and apparently unproblematic but rather crucial step for finding the rest 
is to now define the \emph{modes of the deformed Laplace-Beltrami operator}
as the right hand side of
\begin{eqnarray}
  \label{extdxi coextdxi anticommutator}
  \antiCommutator{\extd_{K,\xi_1}}{\coextd_{K,\xi_2}}
  &=&
  \fatDelta_{K,\xi_1\xi_2}
  \,.
\end{eqnarray}
For this definition to make sense one needs to check that
\begin{eqnarray}
  \label{crucial condition}
  \antiCommutator{\extd_{K,\xi_1\xi_3}}{\coextd_{K,\xi_2}} &=& 
\antiCommutator{\extd_{K,\xi_1}}{\coextd_{K,\xi_2\xi_3}}\,.
\end{eqnarray} It is easy to
verify that this is indeed true for the operators as given in 
\refer{modes of deformed extd coexted on loop space}. 
However, in \S\fullref{deformations} it is found that this condition is a rather strong
constraint on the admissible perturbations of these operators, and the innocent
looking equation \refer{crucial condition} plays a pivotal role
in the algebraic construction of superconformal field theories in the
present context.

With $\fatDelta_{K,\xi}$ consistently defined as in 
\refer{extdxi coextdxi anticommutator}
all remaining brackets follow by using the Jacobi-identity:
\begin{eqnarray}
  \label{remaining brackets}
  \commutator{\frac{1}{2}\fatDelta_{K,\xi_1}}{\extd_{K,\xi_2}}
  &=&
  i\coextd_{K,(\frac{1}{2}\xi_1^\prime\xi_2 - \xi_1\xi_2^\prime)}
  \nonumber\\
  \commutator{\frac{1}{2}\fatDelta_{K,\xi_1}}{\coextd_{K,\xi_2}}
  &=&
  i\extd_{K,(\frac{1}{2}\xi_1^\prime\xi_2 - \xi_1\xi_2^\prime)}
  \nonumber\\
 \nonumber\\
  \commutator{\frac{1}{2}\fatDelta_{K,\xi_1}}{\frac{1}{2}\fatDelta_{K,\xi_2}}
  &=&
  -\, {\cal L}_{K,(\xi_1^\prime\xi_2 - \xi_1\xi_2^\prime)}
  \,.
\end{eqnarray}

In order to make the equivalence to the super-Virasoro algebra
of the algebra thus obtained
more manifest consider the modes of the $K$-deformed Dirac-K{\"a}hler operators 
on loop space:
\begin{eqnarray}
  \label{K-deformed loop space Dirac operator}
  \Dirac_{K,\pm} 
   &\defas& \extd_{K} \pm \coextd_{K}
  \nonumber\\
   &=&
  \Gamma_\mp^{(\mu,\sigma)}\left(\gradOp_{(\mu,\sigma)} \mp iTX^\prime_{(\mu,\sigma)}\right)
  \nonumber\\
  \Dirac_{K,\pm,\xi} &\defas& \extd_{K,\xi} \pm \coextd_{K,\xi}
  \,.
\end{eqnarray}
They are the supercharges which generate the super-Virasoro algebra in the usual chiral form
\begin{eqnarray}
  \label{suVir algebra in almost usual form}
  \antiCommutator{\Dirac_{K,\pm,\xi_1}}{\Dirac_{K,\pm,\xi_2}}
  &=&
  4
  \left(
    \pm
    \frac{1}{2}\fatDelta_{\xi_1\xi_2}
    + 
    i {\cal L}_{\xi_1\xi_2}
  \right)
  \nonumber\\
  \commutator{
    \pm
    \frac{1}{2}\fatDelta_{K,\xi_1}
    + 
    i {\cal L}_{\xi_1}
  }{\Dirac_{K,\pm,\xi_2}}
  &=&
  2 \Dirac_{K,\pm,\frac{1}{2}\xi_1^\prime \xi_2 - \xi_1 \xi_2^\prime}
  \nonumber\\
  \commutator{
    \pm
    \frac{1}{2}\fatDelta_{K,\xi_1}
    + 
    i {\cal L}_{\xi_1}
  }{
    \pm
    \frac{1}{2}\fatDelta_{K,\xi_2}
    + 
    i {\cal L}_{\xi_2}
}
  &=&
  2i
  \left(
    \pm
    \frac{1}{2}\fatDelta_{K,\xi_1^\prime\xi_2 - \xi_1 \xi_2^\prime}
    + 
    i {\cal L}_{\xi_1^\prime\xi_2 - \xi_1 \xi_2^\prime}
  \right)
  \,.
\end{eqnarray}
It is easily seen that this acquires the standard form when we set $\xi\of{\sigma} = e^{in\sigma}$
for $n\in \N$. In order to make the connection with the usual formulation more
transparent consider a flat target space.
If we define the operators
\begin{eqnarray}
  \label{functional definition of Pplusminus}
  {\cal P}_{\pm,(\mu,\sigma)}
  &\defas&
  \frac{1}{\sqrt{2T}}\left(-i\partial_{(\mu,\sigma)} \pm T X^\prime_{(\mu,\sigma)}\right)
\end{eqnarray}  
with commutator
\begin{eqnarray}
  \label{the functional commutator of Pplusminus}
  \commutator
   { {\cal P}_{A,(\mu,\sigma)} }
   { {\cal P}_{B,(\nu,\sigma^\prime)} }
  &=&
  i s_A \delta_{AB}\eta_{\mu\nu}\delta^\prime_{\sigma,\sigma^\prime}
  \,,
  \hspace{1cm}
  \mbox{for}\;g_{\mu\nu} = \eta_{\mu\nu}
\end{eqnarray}
we get, up to a constant factor, the usual modes
\begin{eqnarray}
  \Dirac_{K,\pm,\xi}
  &=&
  \sqrt{2T}i
  \int d\sigma\,
  \xi\of{\sigma}
  \Gamma_{\mp}^\mu\of{\sigma}
  {\cal P}_{\mu,\mp}\of{\sigma}
  \nonumber\\
  \Dirac^2_{K,\pm,\xi^2}
  &=&
  \pm 2T
  \int d\sigma\;
  \left(
  \xi^2\of{\sigma}
  {\cal P}_{\mp}\of{\sigma}\inner {\cal P}_{\mp}\of{\sigma}
  -
  \frac{i}{2}
  \xi\of{\sigma}\left(\xi\Gamma_{\mp}\right)^\prime\of{\sigma}\inner\Gamma_{\mp}\of{\sigma}
  \right)
  \,.
\end{eqnarray}

\subsection{Isomorphisms of the superconformal algebra}
\label{deformations}

The representation of the superconformal algebra as above is manifestly of the
form considered in \S 2.1.1 of \cite{Schreiber:2003a}. We can therefore now
study isomorphisms of the algebra along the lines of \S 2.1.2 of that
paper in order to obtain new SCFTs from known ones. 

From \S 2.1.2 of \cite{Schreiber:2003a} it follows 
that the general continuous isomorphism of the 0-mode sector ($\xi = 1$)
of the algebra \refer{suVir algebra in almost usual form} is induced by some operator
\begin{eqnarray}
  {\bf W} &=& \int d\sigma\; W\of{\sigma}
  \,,
\end{eqnarray}
where $W$ is a field on loop space of unit reparameterization weight 
\begin{eqnarray}
  \label{weight condition on deformation operator}
  w\of{W} = 1\,,
\end{eqnarray}
and looks like
\begin{eqnarray}
  \extd_{K,1} &\mapsto& \extd^{\bf W}_{K,1} \;\defas\; \exp\of{-{\bf W}}\extd_{K,1} \exp\of{{\bf W}}
  \nonumber\\
  \coextd_{K,1} &\mapsto& \coextd^{\bf W}_{K,1} \;\defas\; \exp\of{{\bf W}^\dag}\coextd_{K,1} \exp\of{-{\bf W}^\dag}
  \nonumber\\
  \fatDelta_{K,1} &\mapsto& \fatDelta^{\rm W}_{K,1} \;\defas\; \antiCommutator{\extd^{\bf W}_{K,1}}{\coextd^{\bf W}_{K,1}}
  \nonumber\\
  {\cal L}_1 &\mapsto& {\cal L}_1
  \,.
\end{eqnarray}
This construction immediately generalizes to the full algebra including all modes
\begin{eqnarray}
  \label{isomorphism of Virasoro algebra}
  \extd_{K,\xi} &\mapsto& \extd^{\bf W}_{K,\xi} \;\defas\; \exp\of{-{\bf W}}\extd_{K,\xi} \exp\of{{\bf W}}
  \nonumber\\
  \coextd_{K,\xi} &\mapsto& \coextd^{\bf W}_{K,\xi} \;\defas\; \exp\of{{\bf W}^\dag}\coextd_{K,\xi} \exp\of{-{\bf W}^\dag}
  \nonumber\\
  {\cal L}_\xi &\mapsto& {\cal L}_\xi
\end{eqnarray}
\emph{if}
the crucial relation
\begin{eqnarray}
  \label{crucial relation again}
  \fatDelta^{\bf W}_{K,\xi_1\xi_2} &=& \antiCommutator{\extd^{\bf W}_{K,\xi_1}}{\coextd^{\bf W}_{K,\xi_2}}
\end{eqnarray}
remains well defined, i.e. if \refer{crucial condition} remains true:
\begin{eqnarray}
  \label{crucial condition for deformed operators}
  \antiCommutator{\extd^{\bf W}_{K,\xi_1\xi_3}}{\coextd^{\bf W}_{K,\xi_2}}
  &=&
  \antiCommutator{\extd^{\bf W}_{K,\xi_1}}{\coextd^{\bf W}_{K,\xi_2\xi_3}}
  \,.
\end{eqnarray}

The form of these deformations follows from the fact that no matter which background fields
are turned on, the generator \refer{modes of the reparameterization Lie derivative} of
spatial reparameterizations (at fixed worldsheet time) remains the same, because the
string must be reparameterization invariant in any case. Preservation of the relation
$\mathbf{d}_K^2 = i {\cal L}_K$, which says that $\mathbf{d}_K$ is nilpotent up to
reparameterizations, then implies that $\mathbf{d}_K$ may transform under a similarity
transformation as in the first line of \refer{isomorphism of Virasoro algebra}.  The
rest of \refer{isomorphism of Virasoro algebra} then follows immediately. 

Since this is an important point, at the heart of the approach presented here, 
we should also reformulate it in a more conventional language. Let $L_m$,
$\bar L_m$, $G_m$, $\bar G_m$ be the holomorphic and
antiholomorphic modes of the super Virasoro algebra.
As discussed in \S\fullref{purely gravitational} we have
\begin{eqnarray}
  \label{loop supervirasor in terms of modes}
  \mathbf{\Delta}_{K,\xi} &\propto& L_m + \bar L_{-m}
  \nonumber\\
  \mathbf{\cal L}_{K,\xi} &\propto& L_m - \bar L_{-m}
  \nonumber\\
  \extd_{K,\xi} &\propto& i G_m + \bar G_{-m}
  \nonumber\\
  \coextd_{K,\xi} &\propto& -iG_m + \bar G_{-m} 
  \,,
\end{eqnarray}
with $\xi\of{\sigma} = e^{-im\sigma}$,
as well as
\begin{eqnarray}
  {\mathbf W} &\propto& \sum\limits_n W_n \bar W_n
  \,,
\end{eqnarray}
where $W_m$ and $\bar W_m$ are the modes of the holomorphic and antiholomorphic 
parts of $\mathbf{W}$, which have weight $h$ and $\bar h$, respectively. 
The goal is to find a deformation of \refer{loop supervirasor in terms of modes}
such that $L_m - \bar L_{-m}$ is preserved. Since this is the square of $\pm i G_m + \bar G_{-m}$
the latter may receive a similarity transformation which does not affect $L_m - \bar L_{-m}$
itself. Using $\commutator{L_m}{W_n} = ((h-1)m-n)W_{n+m}$ and similarly for the antiholomorphic part
we see that this is the case for
\begin{eqnarray}
  \label{deformations in mode algebra picture}
  i G_m + \bar G_{-m} &\to& \exp\of{-\sum\limits_n W_n \bar W_n}
  \left(i G_m + \bar G_{-m}\right)\exp\of{\sum\limits_n W_n \bar W_n}
 \nonumber\\
  -i G_m + \bar G_{-m} &\to& \exp\of{\sum\limits_m  \bar W_n^\dagger W_n^\dagger}
  \left(-i G_m + \bar G_{-m}\right)\exp\of{-\sum\limits_n \bar W_n^\dagger W_n^\dagger}
\end{eqnarray}
with 
\begin{eqnarray}
  \label{total unit weight condition on W}
  h + \bar h = 1
  \,,
\end{eqnarray}
because then
\begin{eqnarray}
  L_m - \bar L_{-m} &\to&
  \exp\of{-\sum\limits_n W_n \bar W_n}\left(L_m - \bar L_{-m}\right)
  \exp\of{\sum\limits_n W_n \bar W_n} = L_m - L_{-m}
  \,.
  \nonumber\\
\end{eqnarray}

The point of the loop-space formulation above is to clarify the nature of these deformations,
which in terms of the $L_m, \bar L_m, G_m,\bar G_m$ look somewhat peculiar. In the loop
space formulation it becomes manifest that we are dealing here with a generalization of
the deformations first considered in \cite{Witten:1982} for supersymmetric quantum mechanics, where
the supersymmetry generators are the exterior derivative and coderivative
and are sent by two different similarity transformations
to two new nilpotent supersymmetry generators. This and the relation to the
present approach to superstrings is discussed in detail in section 2.1 of
\cite{Schreiber:2003a}.

Every operator ${\bf W}$ which satisfies \refer{weight condition on deformation operator} 
and \refer{crucial relation again}
hence induces a classical algebra isomorphism of the superconformal algebra
\refer{suVir algebra in almost usual form}. (Quantum corrections to these algebras can be computed
and elimination of quantum anomalies will give background equations of motion, but
this shall not be our concern here.)
Finding such ${\bf W}$ is therefore
a task analogous to finding superconformal Lagrangians in 2 dimensions.

However, two different ${\bf W}$ need
not induce two different isomorphisms.
In particular, \emph{anti-Hermitian} ${\bf W}^\dag = -{\bf W}$ induce 
\emph{pure gauge} transformations in the sense that all algebra elements are
transformed by the \emph{same} unitary similarity transformation
\begin{eqnarray}
  \label{pure gauge deformations}
  {\bf X} &\mapsto& e^{-{\bf W}}{\bf X}e^{\bf W}
  \hspace{1cm}
  \mbox{for ${\bf X}\in \set{\extd_{K,\xi}, \coextd_{K,\xi}, \fatDelta_{K,\xi}, {\cal L}_{\xi}}$
  and ${\bf W}^\dag = - {\bf W}$}
  \,.
\end{eqnarray}
Examples for such unitary transformations are given in 
\S\fullref{A-field background} and \S\fullref{T-duality}.
They are related to background gauge transformations as well as
to string dualities.
For a detailed discussion of the role of such automorphism in the 
general framework
of string duality symmetries see \S 7 of \cite{FroehlichGawedzki:1993}.\\

In the next subsections deformations of the above form are studied in 
general terms and by way of specific examples.

\subsection{NS-NS backgrounds}
\label{NS-NS backgrounds}

We start by deriving superconformal deformations corresponding to 
background fields in the NS-NS sector of the closed Type II string.
Since the conformal weight of an NS-NS vertex comes from a \emph{single} Wick
contraction with the superconformal generators, while that of a spin field, which
enters R-sector vertices, comes from a \emph{double} Wick contraction, the deformation
theory of NS-NS backgrounds is much more transparent than that of NS-R or NS-NS
sectors,  as will be made clear in the following.

\subsubsection{Gravitational background by algebra isomorphism}
\label{gravitational background by algebra isomorphism}

First we reconsider the purely gravitational background from the point of view that
its superconformal algebra derives from
the superconformal algebra for \emph{flat} cartesian target space by a deformation
of the form \refer{isomorphism of Virasoro algebra}. For the point particle
limit this was discussed in equations (38)-(42) of \cite{Schreiber:2003a} and
the generalization to loop space is straightforward:
Denote by
\begin{eqnarray}
  \extd_{K,1}^{\eta}
  &\defas&
  {\cal E}^{\dag (\mu,\sigma)} \partial_{(\mu,\sigma)} + i {\cal E}_{(\mu,\sigma)} X^{\prime(\mu,\sigma)}
\end{eqnarray}
the $K$-deformed exterior derivative on \emph{flat} loop space and define the deformation
operator by
\begin{eqnarray}
  \label{deformation for gravitational background}
  {\bf W} &=& 
  {\cal E}^{\dag} \inner (\ln E) \inner {\cal E}
  \nonumber\\
  &=&
  \int d\sigma\; {\cal E}^{\dag}\of{\sigma} \inner (\ln e\of{X\of{\sigma}}) \inner {\cal E}
  \,,
\end{eqnarray}
where $\ln E$ is the logarithm of a vielbein \refer{vielbein on loop space} on loop space,
regarded as a matrix.
This ${\bf W}$ is constructed so as to satisfy
\begin{eqnarray}
  e^{{\bf W}}{\cal E}^{\dag a}\of{\sigma}e^{-{\bf W}}
  &=&
  \sum_\nu
  e^a{}_\nu
  {\cal E}^{\dag (b=\nu)}
  \,,
\end{eqnarray}
which yields
\begin{eqnarray}
  e^{{\bf W}}
  {\cal E}^{\dag \mu}\of{\sigma}
  e^{-{\bf W}}
  &=&
    e^{{\bf W}}
  e^\mu{}_a {\cal E}^{\dag a}\of{\sigma}
  e^{-{\bf W}}
  \nonumber\\
  &=&
  e^\mu{}_a e^a{}_\nu {\cal E}^{\dag (b=\nu)}
  \nonumber\\
  &=&
  {\cal E}^{\dag (b=\mu)}
  \,.
\end{eqnarray}
Since $e^{\bf W}$ interchanges between two different
vielbein fields which define two different metric tensors the index 
structure becomes a little awkward in the above equations. Since we won't need
these transformations for the further developments we don't bother to 
introduce special notation to deal with this issue more cleanly. The point
here is just to indicate that a $e^{\bf W}$ with the above properties does
exist. It replaces all $p$-forms with respect to $E$ by $p$-forms with respect
to the flat metric. One can easily convince oneself that hence 
the operator $\extd_K$ associated with the
metric $G = E^2$ is related to the operator $\extd_K^\eta$ for flat space by
\begin{eqnarray}
  \label{gravity deformation of extd}
  \extd_{K,\xi}
  &=&
  e^{-{\bf W}}\extd_{K,\xi}^{\eta} e^{{\bf W}}
  \,.
\end{eqnarray}
Therefore, indeed, ${\bf W}$ of \refer{deformation for gravitational background}
induces a gravitational field on the target space.

As was discussed on p. 10 of \cite{Schreiber:2003a} we need to require ${\rm det}\,e = 1$,
and hence 
\begin{eqnarray}
  \label{tracelessness of ln e}
  {\rm tr}\ln e &=& 0
\end{eqnarray}
in order that $\coextd_{K,\xi}^{\bf W} = (\extd_{K,\xi^\ast})^\dag$. This is just a condition
on the nature of the coordinate system with respect to which the metric is constructed in our
framework. As an abstract operator $\extd_{K,\xi}$ is of course
\emph{independent} of any metric, its representation in terms of the operators 
$X^{(\mu,\sigma)},\partial_{(\mu,\sigma)}, {\cal E^{\dag \mu}}, {\cal E}^\mu$ is not, which is what the
above is all about. 

Note furthermore, that
\begin{eqnarray}
  \label{symmetric and antisymmetric ln e}
  {\bf W}^\dag = \pm {\bf W}
  &\Leftrightarrow&
  (\ln e)^{\rm T} = \pm \ln e
  \,.
\end{eqnarray}
According to \refer{pure gauge deformations}
this implies that the antisymmetric part of $\ln e$ generates a pure gauge transformation
and \emph{only the (traceless) symmetric part} of $\ln e$ is responsible for a perturbation of the gravitational 
background. A little reflection shows that the gauge transformation induced by
antisymmetric $\ln e$ is a rotation of the vielbein frame. For further discussion of this
point see pp. 58 of \cite{Schreiber:2001}.

\subsubsection{$B$-field background}
\label{loop space and B-field background}

As in \S 2.1.3 of \cite{Schreiber:2003a} we now consider
the Kalb-Ramond $B$-field 2-form 
\begin{eqnarray}
  B &=& \frac{1}{2}B_{\mu\nu}dx^\mu \wedge dx^\nu
\end{eqnarray}
on target space with field strength $H = dB$. This induces on loop space the 2-form
\begin{eqnarray}
  B_{(\mu,\sigma)(\nu,\sigma^\prime)}\of{X}
  &=&
  B_{\mu\nu}\of{X\of{\sigma}}\delta_{\sigma,\sigma^\prime}
  \,.
\end{eqnarray}
We will study the deformation operator
\begin{eqnarray}
  \label{B-field deformation operator}
  {\bf W}^{(B)}\of{X}
  &\defas&
  \frac{1}{2}
  B_{(\mu,\sigma)(\nu,\sigma^\prime)}\of{X}
  {\cal E}^{\dag (\mu,\sigma)}{\cal E}^{\dag(\nu,\sigma^\prime)}
  \nonumber\\
  &\defas&
  \int d\sigma\;
  \frac{1}{2}B_{\mu\nu}\of{X\of{\sigma}}{\cal E}^{\dag \mu}\of{\sigma}
   {\cal E}^{\dag \nu}\of{\sigma}
\end{eqnarray}
on loop space
(which is manifestly of reparameterization weight 1)
and show that the superconformal algebra that it induces is indeed that found 
by a canonical treatment of the usual supersymmetric $\sigma$-model with 
gravitational and Kalb-Ramond background.  

When calculating the deformations \refer{isomorphism of Virasoro algebra} explicitly
for ${\bf W}$ as in \refer{B-field deformation operator} one finds
\begin{eqnarray}
  \label{b-field deformed loop space susy generators}
  \extd_{K,\xi}^{(B)}
  &\defas&
  \exp\of{-{\bf W}^{(B)}}
  \extd_{K,\xi}
  \exp\of{{\bf W}^{(B)}}
  \nonumber\\
  &=&
  \extd_{K,\xi} + 
  \commutator{\extd_{K,\xi}}{{\bf W}^{(b)}}
  \nonumber\\
  &=&
  \int d\sigma\;
  \xi
  \left(
  {\cal E}^{\dag\mu}\gradOp_\mu
  +
  i T{\cal E}_\mu X^{\prime\mu}
  +
  \frac{1}{6}
  H_{\alpha\beta\gamma}\of{X}
  {\cal E}^{\dag \alpha}{\cal E}^{\dag \beta}
   {\cal E}^{\dag \gamma}
  -
  iT{\cal E}^{\dag\mu}B_{\mu\nu}\of{X}X^{\prime\nu}
  \right)
  \nonumber\\
  \coextd_{K,\xi}^{(B)}
  &=&
  \exp\of{{\bf W}^{\dag(B)}}
  \coextd_K
  \exp\of{-{\bf W}^{\dag(B)}}
  \nonumber\\
  &=&
  -
  \int d\sigma\;
  \xi\of{\sigma}
  \left(
  {\cal E}^{\mu}\gradOp_\mu
  +
  i T{\cal E}^\dag_\mu X^{\prime\mu}
  +
  \frac{1}{6}
  H_{\alpha\beta\gamma}\of{X}
  {\cal E}^{\alpha}{\cal E}^{\beta}
   {\cal E}^{\gamma}
  -
  iT{\cal E}^{\mu}B_{\mu\nu}\of{X}X^{\prime\nu}
  \right)
  \,.
  \nonumber\\
  \,.
\end{eqnarray}
This is essentially equation (72) of \cite{Schreiber:2003a}, with the only difference
that here we have mode functions $\xi$ and an explicit realization of the 
deformation Killing vector.

In order to check that the above is a valid isomorphim condition \refer{crucial condition for deformed operators}
must be calculated. Concentrating on the potentially problematic terms one finds
\begin{eqnarray}
  \antiCommutator{\extd^{(B)}_{K,\xi_1}}{\coextd^{(B)}_{K,\xi_2}}
  &=&
  \int d\sigma\; \xi_1 \xi_2 (\cdots)
  \nonumber\\
  &&+
  \int d\sigma\, d\sigma^\prime
  \xi_1\of{\sigma}\xi_2\of{\sigma^\prime}
    i
    \left(
      {\cal E}^\dag_\mu\of{\sigma^\prime}
      -{\cal E}^\nu\of{\sigma^\prime} B_{\nu\mu}\of{X\of{\sigma^\prime}} 
    \right)
    {\cal E}^{\dag\mu}\of{\sigma}
    \delta^\prime\of{\sigma^\prime,\sigma}
    +
\nonumber\\
  &&+
  \int d\sigma\, d\sigma^\prime
  \xi_1\of{\sigma}\xi_2\of{\sigma^\prime}
    i
    \left(
      {\cal E}_\mu\of{\sigma}
      -{\cal E}^{\dag\nu}\of{\sigma} B_{\nu\mu}\of{X\of{\sigma}} 
    \right)
    {\cal E}^{\mu}\of{\sigma^\prime}
    \delta^\prime\of{\sigma,\sigma^\prime}
  \nonumber\\
  &=&
  \int d\sigma\; \xi_1 \xi_2 (\cdots)
  -i
  \int d\sigma\; 
  \left(
    \xi^\prime_1 \xi_2 
    B_{\nu\mu}{\cal E}^\nu{\cal E}^{\dag \mu}
    +
    \xi_1 \xi_2^\prime
    B_{\nu\mu}{\cal E}^{\dag \nu}{\cal E}^\mu
  \right)
  \nonumber\\
  &=&
  \int d\sigma\; \xi_1 \xi_2 (\cdots)
  \,.
\end{eqnarray}
This expression therefore manifestly satisfies \refer{crucial condition for deformed operators}.

With hindsight this is no surprise, because 
\refer{b-field deformed loop space susy generators} are precisely the 
superconformal generators in functional form as found by canonical analysis of 
the non-linear supersymmetric $\sigma$-model 
\begin{eqnarray}
  \label{susy action of string in b-field background}
  S
  &=&
  \frac{T}{2}
  \int 
    d^2 \xi
    d^2 \theta
    \;
    \left(
      G_{\mu\nu} + B_{\mu\nu}
    \right)
    D_+ {\bf X}^\mu D_- {\bf X}^\nu
    \,,
\end{eqnarray}
where ${\bf X}^\mu$ are worldsheet superfields
\begin{eqnarray}
  {\bf X}^\mu\of{\xi,\theta_+,\theta_-}
  &\defas&
  X^\mu\of{\xi}
  + 
  i\theta_+ \psi_+^{\mu}\of{\xi}
  -
  i \theta_-\psi_-^{\mu}\of{\xi}
  +
  i
  \theta_+
  \theta_- F^\mu\of{\xi}
  \nonumber
\end{eqnarray}
and
$
  D_\pm
  \defas
  \partial_{\theta_{\pm}}
  -i
  \theta_\pm \partial_\pm
$
with
$
  \partial_\pm \defas
  \partial_0 \pm \partial_1
$
are the superderivaties. The calculation can be found in 
section 2 of \cite{Chamseddine:1997}.  
(In order to compare the final result, equations (32),(33)
of \cite{Chamseddine:1997}, with our \refer{b-field deformed loop space susy generators}
note that our fermions $\Gamma_\pm$ are related to the fermions $\psi_\pm$ of
\cite{Chamseddine:1997} by $\Gamma_\pm = (i^{(1\mp 1)/2}\sqrt{2T})\psi_\pm$.)

\subsubsection{Dilaton background}
\label{Dilaton background}
The deformation operator in \refer{deformation for gravitational background} which induces 
the gravitational background 
was of the form ${\bf W} = {\cal E}^\dag \inner M \inner {\cal E}$ with $M$ a traceless
symmetric matrix. If instead we consider a deformation of the same form but for pure trace we get
\begin{eqnarray}
  {\bf W}^{(D)}
  &=&
  -
  \frac{1}{2}
  \int d\sigma\;
  \Phi\of{X}
  {\cal E}^{\dag \mu}{\cal E}_\mu
  \,,
\end{eqnarray}
for some real scalar field $\Phi$ on target space.
This should therefore induce a dilaton background.
The associated superconformal generators are (we suppress the $\sigma$ dependence and the mode
functions $\xi$ from now on)
\begin{eqnarray}
  \label{dilaton deformation}
  \exp\of{-{\bf W}^{(D)}}\extd_{K}\exp\of{{\bf W}^{(D)}}
  &=&
    e^{\Phi/2}{\cal E}^{\dag \mu}
    \left(
      \gradOp_\mu
      -
      \frac{1}{2}(\partial_\mu \Phi){\cal E}^{\dag \nu}{\cal E}_\nu
    \right)
    +iT
    e^{-\Phi/2}
    X^{\prime \mu}
    {\cal E}_\mu
  \nonumber\\
  \exp\of{{\bf W}^{(D)}}\coextd_{K}\exp\of{-{\bf W}^{(D)}}
  &=&
  -
    e^{\Phi/2}{\cal E}^{ \mu}
    \left(
      \gradOp_\mu
      +
      \frac{1}{2}(\partial_\mu \Phi){\cal E}^{\dag \nu}{\cal E}_\nu
    \right)
    -iT
    e^{-\Phi/2}
    X^{\prime \mu}
    {\cal E}^\dag_\mu
  \,.
\end{eqnarray}
It is readily seen that for this deformation equation \refer{crucial condition for deformed operators} 
is satisfied, so that these operators indeed generate a superconformal algebra.

The corresponding Dirac operators are
\begin{eqnarray}
  \label{Phi Dirac operators}
  \extd_K^{(\Phi)}
  \pm 
  \coextd_K^{(\Phi)}
  &=&
  \Gamma_\mp^\mu
  \left(
    e^{\Phi/2}
    \gradOp_\mu
    \mp
    iT
    e^{-\Phi/2}
    G_{\mu\nu}
    X^{\prime\mu}
  \right)
  \mp
  e^{\Phi/2}
  \Gamma_\pm^\mu
  (\partial_\mu \Phi){\cal E}^{\dag \nu}{\cal E}_\nu
  \,.
\end{eqnarray}
Comparison of the superpartners of $\Gamma_{\pm,\mu}$
\begin{eqnarray}
  \mp\frac{1}{2}
  \antiCommutator{  \extd_K^{(\Phi)}
  \pm 
  \coextd_K^{(\Phi)}
  }{
    \Gamma_{\mp,\mu}
  }
  &=&
    e^{\Phi/2}
    \partial_\mu
    \mp
    iT
    e^{-\Phi/2}
    G_{\mu\nu}
    X^{\prime\mu}
  + \mbox{fermionic terms}  
\end{eqnarray}
with equation \refer{Born-Infeld current for pure dilaton background} 
in appendix \S\fullref{Canonical analysis of bosonic D1 brane action} shows that
this has the form expected for the dilaton coupling of a 
D-string.

\subsubsection{Gauge field background}
\label{A-field background}

A gauge field background $A = A_\mu dx^\mu$ should express itself via
$B \to B - \frac{1}{T}F$, where $F = dA$ (e.g. \S 8.7 of 
\cite{  Polchinski:1998}), 
if we assume $A$ to be a $U\of{1}$ connection 
for the moment. Since the present discussion so far refers only
to closed strings and since closed strings have trivial coupling to $A$ it is to be
expected that an $A$-field background manifests itself as a pure gauge 
transformation in the present context. This motivates to investigate the
deformation induced by the anti-Hermitian
\begin{eqnarray}
  \label{A field deformation operator}
  {\bf W}
  &=&
  i A_{(\mu,\sigma)}\of{X}X^{\prime(\mu,\sigma)}
  \;=\;
  i \int d\sigma\;
  A_\mu\of{X\of{\sigma}}X^{\prime\mu}\of{\sigma}
  \,.
\end{eqnarray}
The associated superconformal generators are found to be
\begin{eqnarray}
  \extd_{K}^{(A)(B)}
  &=&
  \extd^{(B)}_{K} +
  i {\cal E}^{\dag \mu}F_{\mu\nu} X^{\prime\nu}
  \nonumber\\
  \coextd_{K}^{(A)(B)}­
  &=&
  \coextd^{(B)}_{K}
  - i {\cal E}^{\mu}F_{\mu\nu} X^{\prime\nu}
  \,.   
\end{eqnarray}
Comparison with \refer{b-field deformed loop space susy generators} shows that indeed
\begin{eqnarray}
  \label{combination of B and F}
  \extd_{K}^{(A)(B)}
  &=&
  \extd_K^{(B-\frac{1}{T}F)}
  \,,
\end{eqnarray}
so that we can identify the background induced by 
\refer{A field deformation operator} with that of the NS 
$U\of{1}$ gauge field.
 
Since
$\exp\of{\bf W}\of{X}$ is nothing but the Wilson loop of $A$ around $X$, it is
natural to conjecture that for a general (non-abelian) gauge field background
$A$ the corresponding deformation is the Wilson loop as well:
\begin{eqnarray}
  \extd_K^{(A)}
  &=&
  \left(
  {\rm Tr}{\cal P}e^{-i \int A_\mu X^{\prime\mu}}
  \right)
  \extd_K
  \left(
  {\rm Tr}{\cal P}e^{+i \int A_\mu X^{\prime\mu}}
  \right)
  \,,
\end{eqnarray}
where $\cal P$ indicates path ordering and $\rm Tr$ the trace in the
Lie algebra, as usual.

\subsubsection{$C$-field background}
\label{C-field background}

So far we have found deformation operators for all massless NS and NS-NS background fields. 
One notes a close similarity between the form of these deformation operators and the
form of the corresponding vertex operators (in fact, the deformation operators are
related to the vertex operators in the (-1,-1) picture. This is discussed in 
\S\fullref{canonical deformations from d to dingens}): 
The deformation operators 
for $G$, $B$ and $\Phi$ are bilinear in the form
creation/annihilation operators on loop space, with the bilinear form (matrix)
seperated into its traceless symmetric, antisymmetric and trace part.

Interestingly, though, there is one more 
deformation operator obtainable by such a bilinear in the form creation/annihilation 
operators. It is
\begin{eqnarray}
  {\bf W}^{(C)} &\defas&
  \frac{1}{2}\int d\sigma\; C_{\mu\nu}\of{X}{\cal E}^{\mu}{\cal E}^\nu
  \,,
\end{eqnarray}
i.e. the \emph{adjoint} of \refer{B-field deformation operator}. It induces the generators
\begin{eqnarray}
  \label{C2 background generators}
  {\extd}^{(C)}_{K,\xi}
  &=&
  \int d\sigma\;
  \xi
  \Big(
    {\cal E}^{\dag \mu}\gradOp_\mu
    +i{\cal E}_\mu X^{\prime \mu}
    -
    {\cal E}^\nu C_\nu{}^\mu \gradOp_\mu
    +
    \frac{1}{2}
    {\cal E}^{\dag \alpha} {\cal E}^\mu {\cal E}^\nu(\nabla_\alpha C_{\mu\nu})
    +
    \nonumber\\
    &&
    -
    \frac{1}{2}
    C_\nu{}^\mu {\cal E}^\nu {\cal E}^\alpha {\cal E}^\beta
      (\nabla_\mu C_{\alpha\beta})
    + \frac{1}{2}C^\alpha{}_\beta{\cal E}^\beta{\cal E}^\mu{\cal E}^\nu(\nabla_\alpha C_{\mu\nu})
  \Big)
  \nonumber\\
  \nonumber\\
  {\coextd}^{(C)}_{K,\xi}
  &=&
  -
  \int d\sigma\;
  \xi
  \Big(
    {\cal E}^{\mu}\gradOp_\mu
    +
    i{\cal E}^\dag_\mu X^{\prime \mu}
    -
    {\cal E}^{\dag \nu} C_\nu{}^\mu \gradOp_\mu
    +
    \frac{1}{2}
    {\cal E}^{\alpha} {\cal E}^{\dag \mu} {\cal E}^{\dag \nu}(\nabla_\alpha C_{\mu\nu})
   \nonumber\\
   &&
   \;\;\;\;
   -\frac{1}{2}
   C_\nu{}^\mu {\cal E}^{\dag \nu}{\cal E}^{\dag\alpha}{\cal E}^{\dag \beta}
    (\nabla_\mu C_{\alpha\beta})
    + \frac{1}{2}C^{ \alpha}{}_\beta{\cal E}^{\dag \beta}{\cal E}^{\dag \mu}{\cal E}^{\dag \nu}
      (\nabla_\alpha C_{\mu\nu})
  \Big)
  \,.
\end{eqnarray}
Furthermore it turns out that this deformation, too, does respect
\refer{crucial condition for deformed operators}:  When we again concentrate only on the potentially
problematic terms we see that
\begin{eqnarray}
  \antiCommutator{{\extd}^{(C)}_{K,\xi_1}}{{\coextd}^{(C)}_{K,\xi_2}}
  &=&
 \int d\sigma \; \xi_1 \xi_2 (\cdots)
  \nonumber\\
  &&
  +
  \antiCommutator{-\int d\sigma\;\xi_1 {\cal E}^\nu C_\nu{}^\mu \gradOp_\mu}
  {-i \int d\sigma\; \xi_2 {\cal E}^{\dag}_\mu X^{\prime \mu}}
  \nonumber\\
  &&
  +
  \antiCommutator{\int d\sigma\;\xi_2 {\cal E}^{\dag \nu} C_\nu{}^\mu \gradOp_\mu}
  {i \int d\sigma\; \xi_1 {\cal E}_\mu X^{\prime \mu}}
  \nonumber\\
  &=&
  \int d\sigma\; \xi_1\xi_2 \left(\cdots\right)
  \nonumber\\
  &&+
  i
  \int d\sigma\;
  \left(
    \xi_1\xi_2^\prime
    \;
    {\cal E}^\dag_\mu C_\nu{}^\mu {\cal E}^\nu 
    +
    \xi_1^\prime \xi_2\;
    {\cal E}_\mu C_\nu{}^\mu {\cal E}^{\dag \nu}
  \right)
  \nonumber\\
  &=&
  \int d\sigma\; \xi_1\xi_2 \left(\cdots\right)  
  \,.
\end{eqnarray}
Therefore \refer{C2 background generators} do generate a superconformal algebra and hence define
an SCFT. 

What, though, is the physical interpretation of the field $C$ on spacetime?
It is apparently not the NS 2-form field, because the generators 
\refer{C2 background generators}
are different from \refer{b-field deformed loop space susy generators}
and don't seem to be unitarily equivalent.
A possible guess would therefore be that it is the \emph{RR 2-form} $C_2$, but
now coupled to a D-string instead of an F-string. 

The description of the F-string in an RR background would involve ghosts and spin fields,
which we do not discuss here. But the coupling of the D-string to the RR 2-form is very
similar to the coupling of the F-string to the Kalb-Ramond 2-form and does not
involve any spin fields. That's why the above deformation might allow an interpretation in 
terms of D-strings in RR 2-form backgrounds.

But this needs to be further examined. A hint in this direction is that
under a duality transformation which changes the sign of the dilaton, the
$C$-field is exchanged with the $B$-field. This is discussed in 
\S\fullref{T-duality for various background fields}.

\subsection{Canonical deformations and vertex operators}
\label{canonical deformations and vertex operators}

With all NS-NS backgrounds under control (\S\fullref{NS-NS backgrounds}) 
we now turn to a more general analysis of the deformations of \S\fullref{deformations}
that puts the results of the previous subsections in perspective and shows
how general backgrounds are to be handled.

\subsubsection{Review of first order canonical CFT deformations}
\label{review of first order canonical CFT deformations}

Investigations of conformal deformations by way of adding terms to the conformal generators
go back as far as\footnote{We are grateful to M. Halpern for making us aware of this work.} 
\cite{FreericksHalpern:1988}, which builds on earlier insights 
\cite{BardakciHalpern:1971,ChodosThorn:1974} into continuous
families of conformal algebras.

It has been noted long ago \cite{OvrutRama:1992} that adding an integrated 
background vertex operator
$V$ (a worldsheet field of weight (1,1))
to the string's action to first order induces a perturbation
\begin{eqnarray}
  L_m &\to&
  L_m + \int d\sigma\; e^{-im\sigma}V\of{\sigma}
\end{eqnarray}
of the Virasoro generators
and a similar shift occurs for the supercurrent \cite{Giannakis:1999}.\\

While in \cite{OvrutRama:1992} this is discussed in CFT language it becomes
quite transparent in canonical language: From the string's worldsheet action
for gravitational $G_{\mu\nu}$, Kalb-Ramond $B_{\mu\nu}$ 
and dilaton $\Phi$ background one finds the classical stress-energy
tensor (\cf \S\fullref{Canonical analysis of bosonic D1 brane action})

\hspace{-2cm}\parbox{1cm}{
\begin{eqnarray}
  T\of{\sigma}
  &=&
  \frac{1}{2}
  G^{\mu\nu}
  \frac{1}{\sqrt{2T}}
  \left(
    e^{\Phi/2}P_\mu + T \left( e^{\Phi/2}B_{\mu\kappa} + e^{-\Phi/2}G_{\mu\kappa} \right)X^{\prime\kappa}
  \right)
  \frac{1}{\sqrt{2T}}
  \left(
    e^{\Phi/2}P_\nu + T \left( e^{\Phi/2}B_{\nu\kappa} + e^{-\Phi/2}G_{\nu\kappa} \right)X^{\prime\kappa}
  \right)\of{\sigma}
  \,,
  \nonumber\\
\end{eqnarray}
}

\hspace{-0.7cm}where $P_\mu$ is the canonical momentum to $X^\mu$.

When expanded in terms of small perturbations
\begin{eqnarray}
  G_{\mu\nu}\of{X}
  &=&
  \eta_{\mu\nu}
  +
  h_{\mu\nu}\of{X} + \cdots
  \nonumber\\
  B_{\mu\nu}\of{X}
  &=&
  0 +
  b_{\mu\nu}\of{X} + \cdots
  \nonumber\\
  \Phi\of{X} &=& 0 + \phi\of{X} + \cdots  
\end{eqnarray}
of the background fields this yields
\begin{eqnarray}
  \label{expanded stress enegery tensor}
  T &\approx&
  \frac{1}{2}
  \left(\eta^{\mu\nu} - h^{\mu\nu}\right)
  \left(
    {\cal P}_{+\mu}
    +
    \sqrt{\frac{T}{2}}
    b_{\mu\kappa}X^{\prime \kappa}
    +
    \sqrt{\frac{T}{2}}
    h_{\mu\kappa}
    X^{\prime \kappa}
    +
    \frac{1}{\sqrt{8T}}\phi P_\mu
    -
    \sqrt{\frac{T}{8}}\phi \eta_{\mu\kappa}X^{\prime\kappa}
  \right)
  \Bigg(
    \cdots
  \Bigg)_\nu
  \nonumber\\
  &=&
  \frac{1}{2}
  \eta_{\mu\nu}
  {\cal P}_{+}^\mu {\cal P}_{+}^\nu
  -
  \underbrace{
  \frac{1}{2}
  h_{\mu\nu}
  {\cal P}_{+}^\mu{\cal P}_{-}^\nu
  }_{\defas V_G}
  -
  \underbrace{
  \frac{1}{2}
  b_{\mu\nu}{\cal P}_+^\mu {\cal P}_-^\nu
  }_{\defas V_B}
  +
  \underbrace{
  \frac{1}{2}
  \phi
  \eta_{\mu\nu}{\cal P}_+^\mu {\cal P}_-^\nu
  }_{\defas V_\Phi}
  +
  \mbox{higher order terms}
  \,,
  \nonumber\\
\end{eqnarray}
where we have defined
\begin{eqnarray}
  {\cal P}^\mu_\pm\of{\sigma}
  &\defas&
  \frac{1}{\sqrt{2T}}
  \left(
    \eta^{\mu\nu}P_\nu \pm T X^{\prime \nu}
  \right)\of{\sigma}\,.
\end{eqnarray}
It must be noted that while the objects ${\cal P}_\pm$, which have Poisson-bracket 
\begin{eqnarray}
  \{{\cal P}^\mu_\pm\of{\sigma},{\cal P}^\nu_\pm\of{\sigma^\prime}\} 
  &=&
  \mp \eta^{\mu\nu}\delta^\prime\of{\sigma-\sigma^\prime}
  \,,
\end{eqnarray}
generate the current algebra of the free theory, they involve, via
$P_\mu = \delta S/\delta \dot X^\mu$, data of the perturbed background and are hence
not proportional to $\partial X$ and $\bar \partial X$.

Still, the first term in \refer{expanded stress enegery tensor} is the generator of the
Virasoro algebra which is associated with the $\mathrm{U}(1)$-currents ${\cal P}_\pm$,
while the following terms are the weight (1,1) vertices 
$V_G$, $V_B$, $V_\Phi$ (with respect to the first term) of
the graviton, 2-form and dilaton, respectively.

Hence in the sense that we regard the canonical coordinates and momenta as fundamental and
hence unaffected by the background perturbation, i.e.
\begin{eqnarray}
  \label{canonical data does not shift}
  X^\mu &\to& X^\mu
  \nonumber\\
  P_\mu &\to& P_\mu
  \,,
\end{eqnarray}
while only the `coupling constants' are shifted
\begin{eqnarray}
  \eta_{\mu\nu} &\to& \eta_{\mu\nu} + h_{\mu\nu}\,,\hspace{1cm}\mbox{etc.}
\end{eqnarray}
we can write
\begin{eqnarray}
  T &\to&
  T + V
  \,,
\end{eqnarray}
where $V$ denotes a collection of weight (1,1) vertices in the above sense.\footnote{
  As is discussed in \cite{OvrutRama:1992}, the 
  issue concerning \refer{canonical data does not shift} in CFT language translates into
the question whether one chooses to treat $\partial X$ and $\bar \partial X$ as free fields
in the perturbed theory and whether the $\partial X \,\partial X$-OPE is taken to receive a
perturbation or not.

For a further discussion of perturbations of SCFTs where this issue is addressed, see
\cite{Schreiber:2003a} and in particular section 2.2.4.
}

CFT deformations of this form are called \emph{canonical deformations} 
\cite{EvansOvrut:1990,EvansGiannakis:1991}.

The central idea of canonical first order deformations is that the (super-) Virasoro algebra
\begin{eqnarray}
  \commutator{T\of{\sigma}}{T\of{\sigma^\prime}}
  &=&
   2i T\of{\sigma^\prime}\delta^\prime\of{\sigma-\sigma^\prime}
   - iT^\prime\of{\sigma^\prime}\delta\of{\sigma-\sigma^\prime}
  +
  A\of{\sigma-\sigma^\prime}
  \nonumber\\
  \antiCommutator{T_F\of{\sigma}}{T_F\of{\sigma^\prime}}
  &=&
  -\frac{1}{2\sqrt{2}}T\of{\sigma^\prime}\delta\of{\sigma^\prime}
  +
  B\of{\sigma-\sigma^\prime}
  \nonumber\\
  \commutator{T\of{\sigma}}{T_F\of{\sigma^\prime}}
  &=&
  \frac{3i}{2}T_F\of{\sigma^\prime}\delta^\prime\of{\sigma-\sigma^\prime}
  -
  i T^\prime_F\of{\sigma^\prime}\delta\of{\sigma-\sigma^\prime}
\end{eqnarray}
(where $A$ and $B$ are the anomaly terms)
together with its chiral partner, generated by $\bar T$ and $\bar T_F$,
is preserved to first order under the perturbation
\begin{eqnarray}
  T\of{\sigma} &\to& T\of{\sigma} + \delta T\of{\sigma}
  \nonumber\\
  T_F\of{\sigma} &\to& T_F\of{\sigma} + \delta T_F\of{\sigma}  
\end{eqnarray}
if, in particular, 
\begin{eqnarray}
  \delta T\of{\sigma} &=& \Phi\of{\sigma}\bar\Phi\of{\sigma}
  \nonumber\\ 
  \delta F\of{\sigma} &=& \Phi_F\of{\sigma}\bar \Phi_F\of{\sigma}
\end{eqnarray}
with
\begin{eqnarray}
  \commutator{T\of{\sigma}}{\Phi\of{\sigma^\prime}}
  &=&
  i \Phi\of{\sigma^\prime}\delta^\prime\of{\sigma-\sigma^\prime} 
  - i 
  \Phi^\prime\of{\sigma^\prime}\delta\of{\sigma-\sigma^\prime}
  \nonumber\\
  \commutator{T\of{\sigma}}{\Phi_F\of{\sigma^\prime}}  
  &=&
  \frac{i}{2} \Phi\of{\sigma^\prime}\delta^\prime\of{\sigma-\sigma^\prime} 
  - i 
  \Phi^\prime\of{\sigma^\prime}\delta\of{\sigma-\sigma^\prime}
  \nonumber\\
  \commutator{T\of{\sigma}}{\bar \Phi\of{\sigma^\prime}}    
  &=& 
  0
  \nonumber\\
  \commutator{T\of{\sigma}}{\bar \Phi_F\of{\sigma^\prime}}    
  &=& 
  0
\end{eqnarray}
and analogous relations for $\delta \bar T$ and $\delta \bar T_F$. 

There are however
also more general fields $\delta T$, $\delta T_F$ of total weight $2$ and $3/2$, 
respectively, which preserve the above super-Virasoro algebra to first order 
\cite{BaggerGiannakis:2001}. But the weight $(1,1)$ part $\Phi\of{\sigma}\bar \Phi\of{\sigma}$
is special in that it corresponds directly to the vertex operator of the background
which is described by the deformed worldsheet theory. 
Further deformation fields of weight different from (1,1) are related to 
\emph{gauge} degrees of 
freedom of the background fields (\cf \cite{BaggerGiannakis:2001} and
the discussion below equation \refer{totally non-gauge delta T}).

\subsubsection{Canonical deformations from $\extd_K \to e^{-\bf W}\extd_K e^{{\bf W}}$.}
\label{canonical deformations from d to dingens}

We would like to see how the deformation theory 
reviewed above
relates to the SCFT deformations that have been
studied in \S\fullref{loop space super-Virasoro generators}.

First recall from \S\fullref{purely gravitational} 
that the chiral bosonic fields in
our notation read
\begin{eqnarray}
  {\cal P}_\pm\of{\sigma}
  &\defas&
  \frac{1}{\sqrt{2T}}
  \left(
    -i
    \frac{\delta}{\delta X}
    \pm
    T X^\prime
  \right)
  \of{\sigma}
\end{eqnarray}
and that according to \S\fullref{exterior algebra over loop space} we
write the worldsheet fermions $\psi$, $\bar \psi$ as $\Gamma_\pm$, respectively, which are
normalized so that
$\antiCommutator{\Gamma^\mu_\pm\of{\sigma}}{\Gamma_\pm^\nu\of{\sigma^\prime}}
= \pm 2 g^{\mu\nu}\of{X\of{\sigma}}\delta\of{\sigma-\sigma^\prime}$, and we
frequently make use of the linear combinations
\begin{eqnarray}
  {\cal E}^{\dag \mu} &=& \frac{1}{2}\left(\Gamma_+^\mu + \Gamma_-^\mu\right)
  \nonumber\\
  {\cal E}^{\mu} &=& \frac{1}{2}\left(\Gamma_+^\mu - \Gamma_-^\mu\right)
  \,.
\end{eqnarray}

In this notation 	the supercurrents for the trivial background read
\begin{eqnarray}
  \label{T_F in terms of d_K}
  T_F\of{\sigma}
  &=&
  \frac{1}{\sqrt{2}}
  \Gamma_+\of{\sigma}
  {\cal P}_+\of{\sigma}
  \;=\;
  \frac{-i}{\sqrt{4T}}
  \left(
    \extd_K\of{\sigma}
    -
    \coextd_K\of{\sigma}
  \right)
  \nonumber\\
  \bar T_F\of{\sigma}
  &=&
  \frac{i}{\sqrt{2}}
  \Gamma_-\of{\sigma}
  {\cal P}_-\of{\sigma}
  \;=\;
  \frac{1}{\sqrt{4T}}
  \left(
    \extd_K\of{\sigma}
    +
    \coextd_K\of{\sigma}
  \right)  
  \,,
\end{eqnarray}
where the $K$-deformed exterior derivative and coderivative on loop space
are identified as
\begin{eqnarray}
  \extd_K &=& 
  \sqrt{T}
  \left(
    \bar T_F
    +
    i
    T_F
  \right)
  \nonumber\\
  \coextd_K &=& 
  \sqrt{T}
  \left(
    \bar T_F
    -
    i T_F
  \right)    
  \,.
\end{eqnarray}
According to \S\fullref{deformations} a consistent deformation of the 
superconformal algebra generated by $T_F$ and $\bar T_F$ is given by
sending
\begin{eqnarray}
  \label{canonical deformation of extd}
  \extd_K\of{\sigma}
  \;\to\;
  \extd^{(W)}_K\of{\sigma}
  &=&
  e^{-{\bf W}}\extd_K\of{\sigma}e^{\bf W}
  =
  \extd_K\of{\sigma}
  +
  \underbrace{
  \commutator{\extd_K\of{\sigma}}{{\bf W}}
  }_{\defas \delta \extd_K\of{\sigma}}
  +
  \cdots
  \nonumber\\
  \coextd_K\of{\sigma}
  \;\to\;
  \coextd^{(W)}_K\of{\sigma}
  &=&
  e^{{\bf W}^\dag}\coextd_K\of{\sigma}e^{-{\bf W}^\dagger}
  =
  \coextd_K\of{\sigma}
  +
  \underbrace{
  \commutator{{\bf W}^\dagger}{\coextd_K\of{\sigma}}
  }_{\defas \delta \coextd_k\of{\sigma}}
  +
  \cdots
\end{eqnarray}
for $\bf W$ some reparameterization invariant operator. From this one finds
$\delta T_F$ by using \refer{T_F in terms of d_K} 
\begin{eqnarray}
  \label{delta T_F and delta bar T_F}
  \delta T_F
  &=&
  -i
  \commutator{\bar T_F}{\frac{1}{2}\left({\bf W} + {\bf W}^\dagger\right)}
  +
  \commutator{T_F}{\frac{1}{2}\left({\bf W} - {\bf W}^\dagger\right)}
  \nonumber\\
  \delta \bar T_F
  &=&
  i
  \commutator{T_F}{\frac{1}{2}\left({\bf W} + {\bf W}^\dagger\right)}
  +
  \commutator{\bar T_F}{\frac{1}{2}\left({\bf W} - {\bf W}^\dagger\right)}  
\end{eqnarray}
which again gives
$\delta T$ by means of 
\begin{eqnarray}
  \label{first order delta T from anticommutators of T_F and delta T_F}
  \antiCommutator{T_F\of{\sigma}}{\delta T_F\of{\sigma^\prime}}
  +
  \antiCommutator{T_F\of{\sigma^\prime}}{\delta T_F\of{\sigma}}
  &=&
  -\frac{1}{2\sqrt{2}}
  \delta T\of{\sigma}
  \delta\of{\sigma-\sigma^\prime}
  \,.
\end{eqnarray}

Before looking at special cases one should note that this necessarily implies that
$\delta T_F$ is of total weight $3/2$ and that $\delta T$ is of total weight $2$. 
That is because $\bf W$, being reparameterization invariant, must be the integral 
(along the string at fixed worldsheet time) over a
field of unit total weight (\cf \refer{weight condition on deformation operator} and
\refer{total unit weight condition on W}) and
because supercommutation with $\extd_K$ or $\coextd_K$ increases the total weight
by $1/2$.

Furthermore, recall from \refer{pure gauge deformations} that the anti-hermitean part
$\frac{1}{2}\left({\bf W} - {\bf W}^\dagger\right)$ of the deformation operator ${\bf W}$
is responsible for \emph{pure gauge transformations} while the hermitean part
$\frac{1}{2}\left({\bf W} + {\bf W}^\dagger\right)$ induces true modifications of the 
background fields. Hence for a pure gauge transformation \refer{delta T_F and delta bar T_F}
yields
\begin{eqnarray}
  \delta T_F &=& \commutator{T_F}{{\bf W}}
  \nonumber\\
  \delta \bar T_F &=& \commutator{\bar T_F}{{\bf W}}
  \,,
  \hspace{1cm}
  \mbox{for ${\bf W}^\dagger = -{\bf W}^\dagger$}
  \,,
\end{eqnarray}
which of course comes from the global similarity transformation \refer{pure gauge deformations}
\begin{eqnarray}
  {\bf X} &\mapsto& e^{-{\bf W}}{\bf X}e^{{\bf W}}\,,
  \hspace{1cm}
  {\bf X} \in \set{T_F, \bar T_F, \cdots}
  \,.
\end{eqnarray}
On the other hand, for a strictly non-gauge transformation 
the transformation \refer{delta T_F and delta bar T_F} simplifies to
\begin{eqnarray}
  \delta T_F
  &=&
  -i
  \commutator{\bar T_F}{{\bf W}}
  \nonumber\\
  \delta \bar T_F
  &=&
  i
  \commutator{T_F}{{\bf W}}
  \,,
  \hspace{1cm}
  \mbox{for ${\bf W}^\dagger = + {\bf W}$}
  \,.
\end{eqnarray}
In the cases where $\bf W$ is antihermitean \emph{and} a $(1/2,1/2)$ field 
(as is in particular the case for the gravitational ${\bf W}^{(G)}$ of 
\S\fullref{gravitational background by algebra isomorphism},
the dilaton ${\bf W}^{(D)}$ of \S\fullref{Dilaton background} 
and the hermitean part of the
Kalb-Ramond ${\bf W}^{(B)} + {\bf W}^{(B)\dagger}$ of 
\S\fullref{loop space and B-field background} )
this, together with \refer{first order delta T from anticommutators of T_F and delta T_F} implies that 
\begin{eqnarray}
  \label{totally non-gauge delta T}
  \delta T
  &\propto&
  \antiCommutator{T_F}{\commutator{\bar T_F}{\bf W}}
\end{eqnarray}
is indeed of weight $(1,1)$, as discussed in the theory of canonical deformations 
\S\fullref{review of first order canonical CFT deformations}.
Furthermore, this shows explicitly that all contributions to $\delta T$ which are of
total weight 2 but \emph{not} of weight (1,1) must come from the antihermitean component 
$\frac{1}{2}\left({\bf W} - {\bf W}^\dagger\right)$ and hence must be associated with background
gauge transformations. (This proves in full generality the respective observation in 
\cite{BaggerGiannakis:2001} concerning 2-form field deformations.)\\

Finally, equation \refer{totally non-gauge delta T} clarifies exactly how the
deformation operators $\bf W$ are related to the \emph{vertex operators} of the
respective background fields, namely it shows that the hermitean part of $\bf W$ is proportional to the
vertex operator in the (-1,-1) picture (i.e. the pre-image under $\antiCommutator{T_F}{\commutator{\bar T_F}{\cdot}}$).\\

As an example, consider the deformation induced by a $B$-field background:

According to \S\fullref{loop space and B-field background} a Kalb-Ramond background is induced by
choosing
\begin{eqnarray}
  {\bf W}
  &=&
  \int d\sigma\;
  \frac{1}{2}
  B_{\mu\nu}\of{X\of{\sigma}}
  {\cal E}^{\dag \mu}
  {\cal E}^{\dag \nu}
\end{eqnarray}
which, using \refer{canonical deformation of extd} gives rise to
\begin{eqnarray}
  \delta \extd_K\of{\sigma}
  &=&
  \left(
  \frac{1}{6}
  H_{\alpha\beta\gamma}\of{X}
  {\cal E}^{\dag \alpha}{\cal E}^{\dag \beta}
   {\cal E}^{\dag \gamma}
  -
  iT{\cal E}^{\dag\mu}B_{\mu\nu}\of{X}X^{\prime\nu}
  \right)
  \of{\sigma}
  \nonumber\\
  \delta\coextd_{K}\of{\sigma}
  &=&
  \left(
    -\frac{1}{6}
    H_{\alpha\beta\gamma}\of{X}
    {\cal E}^{\alpha}{\cal E}^{\beta}
    {\cal E}^{\gamma}
    +
    iT{\cal E}^{\mu}B_{\mu\nu}\of{X}X^{\prime\nu}
  \right)
  \of{\sigma}
\end{eqnarray}
and hence, using \refer{T_F in terms of d_K}, to
\begin{eqnarray}
  \label{B field shift in delta T_F}
  \delta T_F\of{\sigma}
  &=&
  -\frac{i}{12\sqrt{T}}
    H_{\alpha\beta\gamma}\left(
       {\cal E}^{\dag \alpha}{\cal E}^{\dag \beta}{\cal E}^{\dag \gamma}
      +
      {\cal E}^{\alpha}{\cal E}^{\beta}{\cal E}^{\gamma}
    \right)
   -\frac{1}{\sqrt{8}}
   \Gamma_+^\mu
   B_{\mu\nu} \left({\cal P}_+^\nu - {\cal P}_-^\nu\right)
  \,.
\end{eqnarray}
In this special case $\delta T_F$ happens
to be the exact shift of $T_F$ (there are no higher order perturbations of $T_F$ in this background).
As has been noted already in \S\fullref{loop space and B-field background} 
the same result is obtained by canonically
quantizing the supersymmetric $2d$ $\sigma$-model \refer{susy action of string in b-field background} 
which describes superstrings in a Kalb-Ramond background.  

By means of \refer{first order delta T from anticommutators of T_F and delta T_F} the shift
$\delta T$ is easily found to be
\begin{eqnarray}
  \label{delta T for B field background}
  \delta T\of{\sigma}
  &=&
  \Bigg(
    -\frac{1}{12 T}
    \partial_\delta H_{\alpha\beta\gamma}
    \left(
       {\cal E}^{\delta}{\cal E}^{\dag \alpha}{\cal E}^{\dag \beta}{\cal E}^{\dag \gamma}
      +
      {\cal E}^{\dag \delta}{\cal E}^{\alpha}{\cal E}^{\beta}{\cal E}^{\gamma}
   \right)
   -
   i
   \frac{1}{2\sqrt{2T}}
   H_{\alpha\beta\gamma}
    \left(
       {\cal E}^{\dag \alpha}{\cal E}^{\dag \beta}
      +
      {\cal E}^{\alpha}{\cal E}^{\beta}
   \right)       
   {\cal P}_+^\gamma
   \nonumber\\
   &&
   +\frac{i}{\sqrt{4T}}
   \partial_\delta B_{\mu\nu}\left({\cal P}_+^\nu - {\cal P}_-^\nu\right)
   \Gamma_+^\delta\Gamma_+^\mu
   +
   B_{\mu\nu}
   {\cal P}_+^\mu
   {\cal P}_-^\nu
  \Bigg)
  \of{\sigma}
  \,.
\end{eqnarray}
This is of total weight $2$ and contains the weight (1,1)
vertex operator
\begin{eqnarray}
  V &=& B_{\mu\nu}{\cal P}_+^\mu {\cal P}_-^\nu
\end{eqnarray}
of the Kalb-Ramond field
(\cf eqs. (52),(53) in \cite{BaggerGiannakis:2001}). That $T + \delta T$
satisfies the Virasoro algebra to first order at the level of Poisson brackets
follows from the fact that it derives from a consistent deformation of the
form \refer{isomorphism of Virasoro algebra} (as well as from the fact that
it also derives from the respective $\sigma$-model Lagrangian).

\section{Relations between the various superconformal algebras}
\label{relations between susy algebras}

We have found classical deformations of the superconformal algebra associated
with several massless target space background fields. The special algebraic nature
of the form in which we obtain these superconformal algebras admits a convenient
treatment of gauge and duality transformations among the associated background fields.
This is discussed in the following.

\subsection{$\extd_K$-exact deformation operators}
\label{dK exact deformations}

Deformation operators $\bf W$ which are $\extd_K$-exact, i.e. which are of the form
\begin{eqnarray}
  {\bf W}
  &=&
  \superCommutator{\extd_K}{\bf w}
  \,,
\end{eqnarray}
(where $\superCommutator{\cdot}{\cdot}$ is the supercommutator) 
and which furthermore satisfy
\begin{eqnarray}
  \label{condition for extdK-exact W to be algebra preserving}
  {\bf W}
  &=&
  \superCommutator{\extd_{K,\xi}}{{\bf w}_{\xi^{-1}}}  
  \,,\hspace{.7cm}
  \forall\; \xi
\end{eqnarray}
(where ${\bf w}_{\xi} \defas \int d\sigma\; \xi\; w\of{\sigma}$)
are special 
because for them\footnote{
One way to see this is the following:
\begin{eqnarray}
  \superCommutator{\extd_{K,\xi}}{\superCommutator{\extd_K}{\bf w}}
  &=&
  \superCommutator
    {\extd_{K,\xi}}
    {
      \superCommutator
       { \extd_{K,\xi} }
       { {\bf w}_{\xi^{-1}} }
    }
  \nonumber\\
  &=&
  \superCommutator{{\cal L}_{K,\xi^2}}{{\bf w}_{\xi^{-1}}}  
  \nonumber\\
  &=&
  \int d\sigma\; \left(\xi^2 \xi^{-1} w^\prime  + \frac{1}{2}(\xi^2)^\prime\xi^{-1}w\right)\of{\sigma}
  \nonumber\\
  &=&
  \int d\sigma\; \left(\xi w\right)^\prime
  \,,
\end{eqnarray}
where we used that $w\of{\sigma}$ must be of weight $1/2$ in order that
$W\of{\sigma}$ satisfies condition \refer{weight condition on deformation operator}. 
}
\begin{eqnarray}
  \superCommutator{\extd_{K,\xi}}{\superCommutator{\extd_K}{\bf w}}
  &=& 0
\end{eqnarray}
and hence they
leave the generators of the algebra \refer{suVir algebra in almost usual form} 
invariant:
\begin{eqnarray}
  \extd_{K,\xi}^{\bf W}
  &=&
  \extd_{K,\xi}
  \,.
\end{eqnarray}

Two interesting choices for $\bf w$ are
\begin{eqnarray}
  \label{prepotential for AB gauge transformation}
  {\bf w} &=&  A_{(\mu,\sigma)}\of{X} {\cal E}^{\dag {(\mu,\sigma)}}
\end{eqnarray}
and
\begin{eqnarray}
  \label{prepotential for diffeomorphisms}
  {\bf w} &=& V^{(\mu,\sigma)}\of{X} {\cal E}_{{(\mu,\sigma)}}
  \,,
\end{eqnarray}
which both satisfy \refer{condition for extdK-exact W to be algebra preserving}.
They correspond to $B$-field gauge transformations and to diffeomorphisms, respectively:

\subsubsection{$B$-field gauge transformations}

For the choice \refer{prepotential for AB gauge transformation} one gets
\begin{eqnarray}
  {\bf W}
  &=&
  \antiCommutator{\extd_K}{A_{(\mu,\sigma)}{\cal E}^{\dag (\mu,\sigma)}}
  \nonumber\\
  &=&
  \frac{1}{2}(dA)_{(\mu,\sigma)(\nu,\sigma^\prime)}
  {\cal E}^{\dag (\mu,\sigma)}{\cal E}^{\dag (\nu,\sigma^\prime)}
  +
  i T A_{(\mu,\sigma)}X^{\prime (\mu,\sigma)}
  \,.
\end{eqnarray}
Comparison with \refer{B-field deformation operator} and 
\refer{A field deformation operator} shows that this $\bf W$ induces
a $B$-field background with $B = dA$ and a gauge field background
with $F = T \,dA$. According to \refer{combination of B and F} these
two backgrounds indeed precisely cancel.

This ties up a loose end from \S\fullref{loop space and B-field background}: 
A pure gauge transformation $B \to B + dA$ of the $B$-field does not affect
physics of the closed string and hence should manifest itself as an
algebra isomorphism. Indeed, this isomorphism is that induced by
${\bf W} = iT A_{(\mu,\sigma)} X^{\prime (\mu,\sigma)}$.

\subsubsection{Target space diffeomorphisms}

For the choice \refer{prepotential for diffeomorphisms} one gets
\begin{eqnarray}
  {\bf W}
  &=&
  \antiCommutator{\extd_K}{V^{(\mu,\sigma)}{\cal E}_{(\mu,\sigma)}}
  \nonumber\\
  &=&
  \int d\sigma\;
  \left(
    V^{\mu}\partial_{\mu}
    +
    (\partial_\mu V^\nu)
    {\cal E}^{\dag \mu}{\cal E}_{\nu}
  \right)
  \of{\sigma}
  \nonumber\\
  &=&
  {\cal L}_{V}
  \,,
\end{eqnarray}
where ${\cal L}_V$ is the operator inducing the Lie derivative along
$V$ on forms over loop space (\cf A.4 of \cite{Schreiber:2003a}).
According to \S\fullref{gravitational background by algebra isomorphism} 
the part involving
$
    (\partial_\mu V^\nu)
    {\cal E}^{\dag \mu}{\cal E}_{\nu}
$
changes the metric field at every point of target space, while the part involving
$V^\mu\partial_\mu$ translates the fields that enter in the superconformal generators.
This $\bf W$ apparently induces target space diffeomorphisms.

\subsection{T-duality}
\label{T-duality}

It is well known (\cite{LizziSzabo:1998} and references given there)
that in the context of the non-commutative-geometry description of
stringy spacetime physics T-duality corresponds to an inner 
automorphism
\begin{eqnarray}
  {\cal T}: {\cal A} &\to& e^{-{\bf W}}\; {\cal A}\, e^{\bf W} = {\cal A}
\hspace{.7cm}\mbox{with ${\bf W}^\dag = -{\bf W}$}
\end{eqnarray}
of the algebra ${\cal A}$ that enters the spectral triple.
This has been worked out in detail for the bosonic string in
\cite{EvansGiannakis:1996}. In the following this construction
is adapted to and rederived in the present context
for the superstring and then generalized to the various backgrounds
that we have found by deformations.

Following \cite{LizziSzabo:1998}
we first consider T-duality along all dimensions, or
equivalently, restrict attention to the field components along the
directions that are T-dualized. Then we show that 
the \emph{Buscher rules} (see \cite{BandosJulia:2003} for a recent reference) 
for factorized T-duality (i.e. for T-duality along only a single direction)
can very conveniently be derived in our framework, too.

\subsubsection{Ordinary T-duality}

Since T-dualizing along spacetime directions that are not characterized by
commuting isometries is a little subtle (\cf \S4 of \cite{EvansGiannakis:1996}),
assume that a background consisting of a non-trivial metric $g$ and Kalb-Ramond
field $b$ is given together with Killing vectors $\partial_{\mu_n}$
such that
\begin{eqnarray}
  \label{constancy of fields used for T-duality}
  \partial_{\mu_n} g_{\alpha\beta} &=& 0 
  \nonumber\\
  \partial_{\mu_n} b_{\alpha\beta} &=& 0 
  \,.
\end{eqnarray}
For convenience of notation we restrict attention in the following to 
the coordinates $x^{\mu_n}$, since all other coordinates are
mere spectators when T-dualizing. Furthermore we will suppress the
subindex $n$ altogether.

The inner automorphism ${\cal T}$ of the algebra of operators on
sections of the exterior bundle over loop space is defined by
its action on the canonical fields as follows:
\begin{eqnarray}
  \label{T-duality automorphism}
  {\cal T}\of{-i\partial_\mu} &=& X^{\prime \mu}
  \nonumber\\
  {\cal T}\of{X^{\prime \mu}} &=& -i\partial_\mu
  \nonumber\\
  \nonumber\\
  {\cal T}\of{{\cal E}^{\dag a}} &=& {\cal E}_a
  \nonumber\\
  {\cal T}\of{{\cal E}_a} &=& {\cal E}^{\dag a}
  \,.    
\end{eqnarray}
It is possible 
(see \cite{EvansGiannakis:1996}
and pp. 47 of \cite{LizziSzabo:1998}) to express this automorphism manifestly as a
similarity transformation 
\begin{eqnarray}
  {\cal T}\of{A} &=& e^{-\bf W} \,A\, e^{\bf W}
  \,.
\end{eqnarray}
This however requires taking into account normal ordering, 
which would lead us too far afield in the present discussion.
For our purposes it is fully sufficient to
note that ${\cal T}$ preserves the canonical brackets
\begin{eqnarray}
  \label{T duality preserves canonical brackets}
  \commutator{-i\partial_{(\mu,\sigma)}}{X^{\prime (\nu,\sigma^\prime)}} &=& 
   i\delta^\nu_\mu \delta^\prime\of{\sigma,\sigma^\prime}
  \nonumber\\
  &=&
  \commutator{{\cal T}\of{-i\partial_{(i,\sigma)}}}
   {{\cal T}\of{X^{\prime (j,\sigma^\prime)}}}
\end{eqnarray}
and
\begin{eqnarray}
  \antiCommutator{{\cal E}_{(i,\sigma)}}{{\cal E}^{\dag (j,\sigma)}}
  &=&
  \delta_i^j \delta\of{\sigma,\sigma^\prime}
  \nonumber\\
  &=&
  \antiCommutator{{\cal T}\of{{\cal E}_{(i,\sigma)}}}{
    {\cal T}\of{{\cal E}^{\dag (j,\sigma)}}}
\end{eqnarray}
(with the other transformed brackets vanishing)
and must therefore be an algebra automorphism. 

Acting on the $K$-deformed exterior (co)derivative on loop space the transformation
${\cal T}$ produces (we suppress the variable $\sigma$ and the mode functions $\xi$ for
convenience)
\begin{eqnarray}
  {\cal T}\of{
    \extd_{K}
  }
  &=&
  {\cal T}\of{
   {\cal E}^{\dag a}
   E_{a}{}^{\mu}
   \partial_{\mu}
   +
   iT
   {\cal E}_{a}
   E^{a}{}_{\mu}
   X^{\prime \mu}
  }
  \nonumber\\
  &=&
  i{\cal E}_{a}E_{a}{}^{\mu}X^{\prime \mu}
  +
  T {\cal E}^{\dag  a}E^{a}{}_{\mu}\partial_{\mu}  
  \nonumber\\
  &=&
   {\cal E}^{\dag a}
   \tilde E_{a}{}^{\mu}
   \partial_{\mu}
   +
   iT
   {\cal E}_{a}
   \tilde E^{a}{}_{\mu}
   X^{\prime \mu}
  \nonumber\\
  \nonumber\\
  {\cal T}\of{
    \coextd_{K}
  }
  &=&
   \left(
  {\cal T}\of{\extd_{K}}\right)^\dag
  \,,
\end{eqnarray}
where the T-dual vielbein $\tilde E$ is defined as
\begin{eqnarray}
  \tilde E^a{}_\mu &\defas& \frac{1}{T}E_a{}^\mu
  \,.
\end{eqnarray}
(This is obviously not a tensor equation but true in the special coordinates that have been chosen.)
Therefore T-duality sends the deformed exterior derivative associated with the metric defined
by the vielbein $E_a{}^\mu$ to that associated with the metric defined by the vielbein
$\tilde E_a{}^\mu$.
This yields the usual inversion of the spacetime radius $R \mapsto \alpha^\prime/R$:
\begin{eqnarray}
  E^a{}_{\mu} \;=\; \delta^a_\mu\, \sqrt{2\pi} R
  &\Rightarrow&
  \tilde E^a{}_\mu \;=\; \delta^a_\mu\, \frac{1}{T}\frac{1}{\sqrt{2\pi} R} \;=\; 
  \delta^a_\mu \sqrt{2\pi}\; \frac{\alpha^\prime}{R}
  \,.
\end{eqnarray}
Furthermore it is readily checked that the bosonic and fermionic
worldsheet oscillators  
transform as expected:
\begin{eqnarray}
  {\cal T}\of{{\cal P}_{\pm,a}}
  &=&
  {\cal T}\of{
  \frac{1}{\sqrt{2T}}
  \left(
    -i E_a{}^\mu \partial_\mu \pm T E_{a\mu}X^{\prime\mu}
  \right)
  }
  \nonumber\\
  &=&
  \frac{1}{\sqrt{2T}}
  \left(
     E_a{}^\mu X^{\prime\mu} \pm -iT E_{a\mu}\partial_\mu
  \right)
  \nonumber\\
  &=&
  \pm 
  \frac{1}{\sqrt{2T}}
  \left(-i
    \tilde E_a{}^\mu \partial_\mu
    \pm 
    T 
    E_{a\mu}X^{\prime\mu	}
  \right)
  \nonumber\\
  &=&
  \pm \tilde {\cal P}_{\pm,a}
\end{eqnarray}
and
\begin{eqnarray}
  {\cal T}\of{\Gamma^a_\pm}
  &=&
  \pm \Gamma^a_\pm
  \,.
\end{eqnarray}

More generally, when the Kalb-Ramond field is included one finds
\begin{eqnarray}
  {\cal T}\of{
  \extd_{K}^{(B)}
  \pm
  \coextd_{K}^{(B)}
  }
  &=&
  {\cal T}\of{
    \Gamma_\mp^a
    E_a{}^\mu
  \left(
    \partial_\mu
    \mp
    iT
    \left(
      G_{\mu\nu} \pm
      B_{\mu\nu}  
    \right)
   X^{\prime \nu}
  \right)
  }
  \nonumber\\
  &=&
  \mp
    \Gamma_\mp^a
    E_a{}^\mu
  \left(
    i X^{\prime\mu }
    \mp
    T
    \left(
      G_{\mu\nu}
      \pm
      B_{\mu\nu}  
    \right)
    \partial_\nu
  \right)
  \nonumber\\
  &=&
    \Gamma_\mp^a
    \tilde E_a{}^\mu
  \left(
    \partial_\mu
    \mp
    [T
    \left(
      G_{\mu\nu} \pm B_{\mu\nu}  
    \right)]^{-1}
    X^{\prime \nu}
  \right)
\end{eqnarray}
with
\begin{eqnarray}
  \tilde E_a{}^\mu
  &\defas&
  T
  E_a{}^\nu (G_{\nu\mu}\pm B_{\nu\mu})
  \,,
\end{eqnarray}
which reproduces the well known result 
(equation (2.4.39) of \cite{GiveonPorratiRabinovici:1994})
that the T-dual spacetime metric is given by
\begin{eqnarray}
  \tilde G^{\mu\nu}
  &=&
  T^2
  [(G\mp B)G^{-1}(G \pm B)]_{\mu\nu}
\end{eqnarray}
and that the T-dual Kalb-Ramond field is
\begin{eqnarray}
  \tilde B_{\mu\nu} &=& \pm[\frac{1}{T^2}(G\pm B)^{-1} -\tilde G]_{\mu\nu}
  \nonumber\\
  &=&
  \left[
    T^2(G\mp B)B^{-1}(G\pm B)
  \right]^{-1}_{\mu\nu}
  \,.
\end{eqnarray}
\begin{eqnarray}
\end{eqnarray}
It is also very easy in our framework to derive the Buscher rules 
for T-duality along a single direction $y$ (``factorized duality''): 
Let $\mathcal{T}_y$ be the transformation \refer{T-duality automorphism} restricted to
the $\partial_y$ direction, then 
from
\begin{eqnarray}
  \label{Buscher transformation}
  {\cal T}\of{
  \extd_{K}^{(B)}
  \pm
  \coextd_{K}^{(B)}
  }
  &=&
    \Gamma_\mp^a
    \left(
    E_a{}^i
    \partial_i
    \mp
    iT
    E_a{}^\mu
    \left(
      G_{\mu i} \pm
      B_{\mu i}  
    \right)
   X^{\prime i}
   \right)
  \nonumber\\
   &&+
  {\cal T}\of{
    \Gamma_\mp^a
    \left(
    E_a{}^y
    \partial_y
    \mp
    iT
    E_{}^\mu
    \left(
      G_{\mu y} \pm
      B_{\mu y}  
    \right)
   X^{\prime y}
  \right)
  }
  \nonumber\\
  &=&
    \Gamma_\mp^a
    \left(
    E_a{}^i
    \partial_i
    \mp
    iT
    E_a{}^\mu
    \left(
      G_{\mu i} \pm
      B_{\mu i}  
    \right)
   X^{\prime i}
   \right)
  \nonumber\\
   &&
  +
    \Gamma_\mp^a
    \left(
    T
    E_a{}^\mu
    \left(
      G_{\mu y} \pm
      B_{\mu y}  
    \right)
   \partial_y
   \mp
    i
    E_a{}^y
    e^{\Phi/2}
    X^{\prime y}
  \right)
\end{eqnarray}
one reads off the T-dual inverse vielbein
\begin{eqnarray}
  \tilde E_a{}^i &=& E_a{}^i
  \nonumber\\
  \tilde E_a{}^y &=& 
      T
    E_a{}^\mu
    \left(
      G_{\mu y} \pm
      B_{\mu y}  
    \right)
\end{eqnarray}
whose inverse $\tilde E_\mu{}^a$ is easily seen to be
\begin{eqnarray}
  \tilde E_{i}{}^a
  &=&
  E_i{}^a -\frac{G_{iy}\pm B_{iy}}{G_{yy}}E_y{}^a
  \nonumber\\
  \tilde E_y{}^a 
  &=&
  \frac{1}{T G_{yy}}E_y{}^a
  \,,
\end{eqnarray}
which gives the T-dual metric with minimal computational effort:
\begin{eqnarray}
  \tilde G_{yy}
  &=&
  \frac{1}{T G_{yy}}
  \nonumber\\
  \tilde G_{iy}
  &=&
  \mp \frac{B_{iy}}{TG_{yy}}
  \nonumber\\
  \tilde G_{ij}
  &=&
    G_{ij} - \frac{1}{G_{yy}}
    \left(
      G_{iy}G_{jy}
      -
      B_{iy}B_jy
    \right)
  \,.
\end{eqnarray}
Similarly the relations 
\begin{eqnarray}
  \tilde E_a{}^\mu(\tilde G_{\mu i}\pm \tilde B_{\mu i})
  &=&
  E_a{}^\mu(G_{\mu i} \pm B_{\mu i})
  \nonumber\\
  \tilde E_a{}^\mu(\tilde G_{\mu y}\pm \tilde B_{\mu y})
  &=&
  \frac{1}{T}
  E_a{}^y
\end{eqnarray}
for the T-dual $B$-field $\tilde B$ are read
off from \refer{Buscher transformation}. Solving them for $\tilde B$ is
straightforward and yields
\begin{eqnarray}
  \tilde B_{ij}
  &=&
  B_{ij}
  \mp
  \frac{1}{G_{yy}}
  \left(
    B_{jy}G_{iy} - B_{iy}G_{iy}
  \right)
  \nonumber\\
  \tilde B_{iy}
  &=&
  \frac{1}{T G_{yy}}G_{iy}
  \,.
\end{eqnarray}
These are the well known \emph{Buscher rules} for factorized T-duality
(see eq. (4.1.9) of \cite{GiveonPorratiRabinovici:1994}).

The constant dilaton can be formally absorbed into the string tension $T$ 
and is hence seen to be invariant under ${\cal T}_y$. This is
correct in the classical limit that we are working in. It
is well known (e.g. eq. (4.1.10) of \cite{GiveonPorratiRabinovici:1994}),
that there are higher loop corrections to the T-dual dilaton. These corrections are 
not visible with the methods discussed here.

$\,$\\

Using our representation for the superconformal generators
in various backgrounds it is now straightfroward to include more general background fields than
just $G$ and $B$ in the above construction:

\subsubsection{T-duality for various backgrounds}
\label{T-duality for various background fields}

When turning on all fields $G$, $B$, $A$, $C$ and $\Phi$, requiring them to be
constant in the sense of \refer{constancy of fields used for T-duality} and assuming
for convenience of notation that $B\inner C = 0$ the supercurrents read according
to the considerations in \S\ref{purely gravitational}-\S\ref{A-field background}
\begin{eqnarray}
  \label{general supercurrents for constant fields}
    \extd_K^{(\Phi)(A+B+C)} \pm \coextd_K^{(\Phi)(A+B+C)}
  &=&
    \Gamma_\mp^a E_a{}^\mu
    \left(
       e^{\Phi/2}(G_\mu{}^\nu \pm C_\mu{}^\nu)\partial_\nu
       \mp iT e^{-\Phi/2}(G_{\mu\nu}\pm (B_{\mu\nu}+\frac{1}{T}F_{\mu\nu})X^{\prime \nu}
    \right)
  \,.
  \nonumber\\
\end{eqnarray}

It is straightforward to apply $\cal T$ to this expression and read off the
new fields. However, since the resulting expressions are not too
enlightening we instead use a modification $\tilde {\cal T}$ of $\cal T$, which, too,
induces an algebra isomorphism, but which produces more
accessible field redefinitions. The operation $\tilde {\cal T}$ differs
from $\cal T$ in that index shifts are included:
\begin{eqnarray}
  \tilde {\cal T}\of{-i\partial_\mu}
  &\defas&
  T
  g_{\mu\nu}X^{\prime \nu}
  \nonumber\\
  \tilde {\cal T}\of{X^{\prime\mu}}
  &\defas&
  -\frac{i}{T}g^{\mu\nu}\partial_\nu
  \nonumber\\
  \nonumber\\
  \tilde {\cal T}\of{{\cal E}^{\dag a}}
  &\defas&
  {\cal E}^a
  \nonumber\\
  \tilde {\cal T}\of{{\cal E}^{a}}
  &\defas&
  {\cal E}^{\dag a}  
  \,.
\end{eqnarray}
Due to the constancy of $g_{\mu\nu}$ this preserves the canonical brackets
just as in \refer{T duality preserves canonical brackets} and hence is
indeed an algebra isomorphism.

Applying it to the supercurrents \refer{general supercurrents for constant fields} yields
\begin{eqnarray}
  \tilde {\cal T}
  \left[
    \extd_K^{(\Phi)(B+C)} \pm \extd_K^{(\Phi)(B+C)}
  \right]
  &=&
    \Gamma_\mp^a E_a{}^\mu
    \left(
        e^{-\Phi/2}(G_{\mu}{}^{\nu}\pm (B_{\mu}{}^{\nu} + \frac{1}{T}F_\mu{}^\nu)\partial_\nu
        \mp i Te^{\Phi/2}(G_{\mu\nu} \pm C_{\mu\nu}) X^{\prime \nu}
    \right)  
  \,.
  \nonumber\\
\end{eqnarray}
Comparison shows that under $\tilde {\cal T}$ the background fields transform as
\begin{eqnarray}
  \label{possible S-duality}
  B_{\mu\nu}+ \frac{1}{T}F_{\mu\nu} &\to& C_{\mu\nu}
  \nonumber\\
  C_{\mu\nu} &\to& B_{\mu\nu} + \frac{1}{T}F_{\mu\nu}
  \nonumber\\
  G_{\mu\nu} &\to& G_{\mu\nu}
  \nonumber\\
  \Phi &\to& -\Phi
  \,.
\end{eqnarray}
The fact that under this transformation 
the NS-NS 2-form is exchanged with what we interpreted as the 
R-R 2-form and that the
dilaton reverses its sign is reminiscent of S-duality. 
It is well known \cite{Duff:1995} that T-duality and S-duality are
themselves dual under the exchange of the fundamental F-string and the 
D-string. How exactly this applies to the constructions
presented here remains to be investigated. (For instance the
sign that distinguishes \refer{possible S-duality} from
the expected result would need to be explained, maybe by a change
of orientation of the string.)

\subsection{Hodge duality on loop space}
\label{Hodge duality on loop space}

For the sake of completeness in the following the relation of
loop space Hodge duality to the above discussion is briefly indicated.
It is found that ordinary Hodge duality is at least superficially related to
the algebra isomorphisms discussed in \S\fullref{T-duality}. 
Furthermore a deformed version of Hodge duality is considered which
preserves the familiar relation $\extd^\dagger = \pm \star\,\extd\,\star^{-1}$.

\subsubsection{Ordinary Hodge duality}

On a finite dimensional pseudo-Riemannian manifold, let $\bar \star$ be the
phase-shifted Hodge star operator which is normalized so
as to satisfy
\begin{eqnarray}
  (\bar \star)^\dag &=& - \bar \star
  \nonumber\\
  (\bar \star)^2 &=& 1
  \,.
\end{eqnarray}
(For the precise relation of $\bar \star$ to the ordinary Hodge $\star$ see (A.18) of \cite{Schreiber:2003a}.)
The crucial property of this operator can be expressed as
\begin{eqnarray}
  \bar \star\, \onbCreator^{\mu} &=& \onbAnnihilator^\mu\, \bar \star
  \,,
\end{eqnarray}
where $\onbCreator^\mu$ is the operator of exterior multiplication by $dx^\mu$
and $\onbAnnihilator^\mu$ is its adjoint under the Hodge inner product. 

It has been pointed out in \cite{Witten:1982} 
that the notion of Hodge duality can be carried over to
infinite dimensional manifolds. This means in particular that on loop space there
is an idempotent operator $\bar\star$ so that
\begin{eqnarray}
  \bar \star\, {\cal E}^{\dag \mu} &=& {\cal E}^\mu\, \bar \star
\end{eqnarray}
and
\begin{eqnarray}
  \commutator{\bar \star}{X^{(\mu,\sigma)}}
  = 0 = 
  \commutator{\bar \star}{\gradOp_{(\mu,\sigma)}}
  \,. 
\end{eqnarray}
It follows in particular that the $K$-deformed exterior derivative is
related to its adjoint by
\begin{eqnarray}
  \coextd_K &=& - \bar \star\; \extd_K \; \bar \star
  \,.
\end{eqnarray}
In fact this holds for all the modes:
\begin{eqnarray}
  \coextd_{K,\xi}^\dag &=& - \bar \star\; \extd_{K,\xi} \; \bar \star
  \,.
\end{eqnarray}

In the spirit of the discussion of T-duality by algebra isomorphisms in 
\S\fullref{T-duality} one can equivalently say that $\bar \star$ induces
an algebra isomorphism $\cal H$ defined by
\begin{eqnarray}
  {\cal H}\of{A} &\defas& \bar \star\; A \bar \star
  \,,
\end{eqnarray}
i.e.
\begin{eqnarray}
  {\cal H }\of{-i\partial_\mu} &=& -i\partial_\mu
  \nonumber\\
  {\cal H}\of{X^\mu} &=& X^\mu
  \nonumber\\
  \nonumber\\
  {\cal H}\of{{\cal E}^{\dag a}} &=& {\cal E}^{a}
  \nonumber\\
  {\cal H}\of{{\cal E}^{a}} &=& {\cal E}^{\dag a}
  \,.  
\end{eqnarray}

It is somewhat interesting to consider the result of first applying $\cal H$ to
$\extd_K$ and then acting with the deformation operators $\exp\of{\bf W}$
considered before. This is equivalent to considering the deformation obtained by
$\bar \star\, e^{\bf W}$. This yields
\begin{eqnarray}
  \extd_K
  &\to&
  \left(e^{-\bf W}\bar \star\right)
  \extd_K
  \left(
  \bar \star\, e^{\bf W}\right)
  \;=\;
  -
  e^{-\bf W}\,\coextd_K \,e^{\bf W}
  \nonumber\\
  \coextd_K
  &\to&
  \left(e^{{\bf W}^\dag}\bar \star\right)
  \coextd_K
  \left(
  \bar \star\, e^{-{\bf W}^\dag}\right)
  \;=\;
  -
  e^{{\bf W}^\dag}\,\extd_K \,e^{-{\bf W}^\dag}
  \,.
\end{eqnarray}
Hence, except for a global and irrelevant sign, the deformations induced by
$e^{\bf W}$ and $\bar \star\, e^{\bf W}$ are related by
\begin{eqnarray}
  {\bf W} &\leftrightarrow& -{\bf W}^\dag
  \,.
\end{eqnarray}
Looking back at the above results for the backgrounds induced by various ${\bf W}$ 
this corresponds to
\begin{eqnarray}
  B &\leftrightarrow& C
  \nonumber\\
  A &\leftrightarrow& A
  \nonumber\\
  \Phi &\leftrightarrow& -\Phi
  \,.
\end{eqnarray}

It should be noted though, that unlike the similar correspondence 
\refer{possible S-duality} both sides of this relation are not unitarily
equivalent in the sense that the corresponding superconformal generators
$e^{-\bf W}\bar \star\, \extd_K \, \bar \star e^{\bf W}$
and
$e^{-\bf W} \, \extd_K \,  e^{\bf W}$
are not unitarily equivalent.

Nevertheless, it might be that the physics described by both generators
is somehow related. This remains to be investigated.

\subsubsection{Deformed Hodge duality}
\label{deformed Hodge duality}

The above shows that for general background fields (general deformations of the
superconformal algebra) the familiar equality of
$\coextd^{\bf W}_{K,\xi}$ with $- {\bar \star} \,\extd^{\bf W}_{K,\xi}\,{\bar \star}^{-1}$
is violated. It is however possible to consider a deformation ${\bar\star}^{\bf W}$ 
of $\bar \star$ itself
which restores this relation:
\begin{eqnarray}
  {\bar \star}^{\bf W}
  &\defas&
  e^{{\bf W}^\dag}\,\bar\star\, e^{\bf W}
  \,.
\end{eqnarray}
Obviously this operator satisfies
\begin{eqnarray}
  \coextd_{K,\xi}^{\bf W}
  &=&
  - {\bar \star}^{\bf W}
  \,
    \extd_{K,\xi}
  \,
  ({\bar \star}^{\bf W})^{-1}	
  \,.
\end{eqnarray}

The Hodge star remains invariant under this deformation when
$\bf W$ is anti-Hodge-dual:
\begin{eqnarray}
  \bar \star = {\bar \star}^{\bf W}
  &\Leftrightarrow&
  \bar \star \,{\bf W}\,\bar \star
  =
  - {\bf W}^\dagger
  \,.
\end{eqnarray} 
This is in particular true for the gravitational deformation of
\S\fullref{gravitational background by algebra isomorphism}.
It follows that ${\bar \star}^{(G)} = {\bar \star}$. This
can be understood in terms of the fact that the definition of the
Hodge star involves only the orthonormal metric on the tangent space
(\cf (A.14) of \cite{Schreiber:2003a}).

\subsection{Deformed inner products on loop space.}
\label{deformed inner products on loop space}

The above discussion of deformed Hodge duality on loop space motivates the
following possibly interesting observation:

From the point of view of differential geometry the exterior derivative
$\extd$ on a manifold is a purely topological object which does not
depend in any way on the geometry, i.e. on the metric tensor. The geometric information
is instead contained in the Hodge star operator $\star$, the 
Hodge inner product  
$\langle \alpha |\beta\rangle = \int\limits_\manifold \alpha \wedge \star \beta$
on differential forms
and the adjoint $\coextd$ of $\extd$ with respect to $\langle\cdot|\cdot\rangle$.

We have seen in \S\fullref{deformed Hodge duality} that deformations of the
Hodge star operator on loop space may encode not only information about the
geometry of target space, but also about other background fields, like Kalb-Ramond
and dilaton fields. These deformations are accompanied by analogous deformations
\refer{isomorphism of Virasoro algebra} of $\extd$ and $\coextd$.

But from this point of view of differential geometry it appears unnatural to associate a 
deformation of both $\coextd$ as well as $\extd$ with a deformed Hodge star operator.
One would rather expect that $\extd$ remains unaffected by any background fields while
the information about these is contained in $\star$, $\langle\cdot|\cdot \rangle$ and 
$\coextd.$

Here we want to point out that both points of view are equivalent and indeed
related by a global similarity transformation ('duality') and that the change in 
point of view makes an interesting relation to noncommutative geometry transparent.

Namely consider deformed operators
\begin{eqnarray}
  \label{deformed ops once again}
  \extd^{({\bf W})}
  &=&
  e^{-{\bf W}}\extd e^{\bf W}
  \nonumber\\
  \coextd^{({\bf W})}
  &=&
  e^{{\bf W}^\dagger}\coextd e^{-{\bf W}^\dagger}
\end{eqnarray}
on an inner product space $\cal H$ with inner product $\langle\cdot | \cdot \rangle$
as in \refer{isomorphism of Virasoro algebra}.

By applying a global similarity transformation
\begin{eqnarray}
  \ket{\psi} &\to& \tilde {\ket{ \psi}} \defas e^{{\bf W}}\ket{\psi}
  \nonumber\\
  A &\to& \tilde A \defas e^{\bf W}Ae^{- {\bf W}}
\end{eqnarray}
to all elements $\ket{\psi} \in {\cal H}$ and all operators $A$ on $\cal H$
one of course finds
\begin{eqnarray}
  \left({\extd^{(\bf W)}}\right)^{\tilde{}}
  &=&
  \extd
  \nonumber\\
  \left({\coextd^{({\bf W})}}\right)^{\tilde {}}
  &=&
  e^{{\bf W} + {\bf W}^\dagger}\coextd e^{-{\bf W}- {\bf W}^\dagger}
  \,.
\end{eqnarray}

By construction, the algebra of $\left({\extd^{(\bf W)}}\right)^{\tilde{}}$
and $\left({\coextd^{(\bf W)}}\right)^{\tilde{}}$ is the same as that of
${\extd^{(\bf W)}}$ and ${\coextd^{(\bf W)}}$. But now all the information about the
deformation induced by $\bf W$ is contained in $\left({\coextd^{(\bf W)}}\right)^{\tilde{}}$
alone. This has the advantage that we can consider a deformed innner product
\begin{eqnarray}
  \label{deformed inner product}
  \langle\cdot | \cdot \rangle_{{}_{({ \bf W})}}
  &\defas&
  \langle\cdot | \,e^{-({\bf W}+{\bf W}^\dagger)}\, \cdot \rangle  
\end{eqnarray}
on ${\cal H}$ with respect to which 
\begin{eqnarray}
  \extd^{\dagger_{(\bf W)}}
  &=&
 \left({\coextd^{(\bf W)}}\right)^{\tilde{}}
  \,,
\end{eqnarray}
where $\langle A \cdot | \cdot \rangle_{{}_{({ \bf W})}} \defas 
\langle\cdot | A^{\dagger_{({\bf W})}} |\cdot \rangle_{{}_{({ \bf W})}}
$. If the original inner product came from a Hodge star this corresponds to
a deformation
\begin{eqnarray}
  \star &\to& \star \;e^{-({\bf W} + {\bf W}^\dagger)}
 \,.
\end{eqnarray}
This way now indeed the entire deformation comes from a deformation of the Hodge star
and the inner product.

That this is equivalent to the original notion \refer{deformed ops once again} 
of deformation can be checked again by noting that the deformed inner product
of the deformed states agrees with the original inner prodcut on the original states
\begin{eqnarray}
  \langle \tilde \psi | \tilde \phi \rangle_{{}_{({\bf W})}}
  &=&
  \langle \psi | \phi \rangle 
  \,.
\end{eqnarray}

These algebraic manipulations by themselves are rather trivial, but the interesting aspect
is that the form \refer{deformed inner product} of the deformation appears
in the context of noncommutative spectral geometry \cite{ForgySchreiber:2004}.
The picture that emerges is roughly that of a spectral triple 
$(\mathcal{A},\extd_K \pm \extd^{\dag_{({\mathbf W})}}_K, \mathcal{H})$,
where $\mathcal{A}$ is an algebra of functions on loop space 
(\cf \fullref{loop space definitions}), $\mathcal{H}$ is the inner product
space of differential forms over loop space equipped with a deformed Hodge
inner product \refer{deformed inner product} which encodes all the information
of the background fields on target space, and where two Dirac operators are given by
$\extd_K \pm \extd^{\dag_{({\mathbf W})}}_K$. There have once been attempts
\cite{Chamseddine:1997,Chamseddine:1997b,FroehlichGrandjeanRecknagel:1997,
FroehlichGrandjeanRecknagel:1996,FroehlichGawedzki:1993} 
to understand the superstring by regarding the worldsheet supercharges
as Dirac operators in a spectral triple. Maybe the insight that 
and how target space background fields manifest themselves as simple algebraic
deformations \refer{isomorphism of Virasoro algebra} of the Dirac operators, or, equivalently, 
\refer{deformed inner product} of the inner product on $\mathcal{H}$ can help to make progress with this approach.

\newpage
\section{Summary and Discussion}
\label{summary and discussion}

We have noted that
the loop space formulation of the superstring highlights its 
Dirac-K{\"a}hler structure that again emphasizes the role played by
the linear combinations of the leftmoving supercurrent $G$ and its rightmoving 
counterpart $\tilde G$,
which are thus seen to be generalized exterior derivative 
$\extd_K$ and coderivative $\coextd_K$
on loop space. This fact led us to the study of deformations
$\extd_K \to e^{-\bf W}\,\extd_K\, e^{\bf W}$ which preserve the
superconformal algebra at the level of Poisson brackets. 
One important point is that under these deformations the
usual supercurrents $G$ and $\tilde G$ transform as
\begin{eqnarray}
  \label{deformation in conclusion}
  \left.
 \begin{array}{c}
   G\\ \tilde G
 \end{array}
  \right\rbrace  
 \propto \extd_K \pm \coextd_K &\to&
  e^{-\bf W}\,\extd_K\, e^{\bf W}
  \pm
  e^{{\bf W}^\dag}\,\coextd_K\, e^{-{\bf W}^\dag}
  \,,
 \nonumber
\end{eqnarray}
so as to preserve the generator of reparameterizations of the string at fixed 
worldsheet time.

We have shown that the above deformations reproduce, when truncated at first order,
the well-known canonical deformations. The hermitean part of the deformation operator $\bf W$
was found to be the vertex operator of the respective background in the (-1,-1) superghost
picture and the anti-hermitean part was seen to give rise to gauge transformations of the
background fields.

In the loop-space notation this means that for the NS-NS and NS fields one finds the
following list of deformation operators:
\begin{center}
\begin{tabular}{|l|l|c|}
  \hline
  background field & & deformation operator $\bf W$\\
  \hline\hline
  metric & $G = E^2$   &  ${\cal E}^{\dag} \inner (\ln E) \inner {\cal E}$\\
  \hline
  Kalb-Ramond & $B$ &  $\frac{1}{2}{\cal E}^\dag \inner B \inner {\cal E}^\dag$\\
  \hline
  dilaton          & $\Phi$ &  $-\frac{1}{2} \Phi {\cal E}^\dag \inner {\cal E}$ \\
  \hline
  2-form for D-string (?)      & $C$    &  $\frac{1}{2} {\cal E} \inner C \inner {\cal E}$\\
  \hline
  gauge connection & $A$    &  $ i A \inner X^\prime$\\
  \hline
\end{tabular}
\end{center}
where ${\cal E}^\dag$ and $\cal E$ are form creation/annihilation operators on loop space.

Using these deformations the explit (functional/canonical) form of the superconformal
generators for all these backgrounds has been obtained, which allowed the 
study of T-duality by means of algebra isomorphisms. It turned out that under
a certain modification of ordinary T-duality the background fields transform as
\begin{eqnarray}
  C &\leftrightarrow& B + \frac{1}{T}F
  \nonumber\\
  G &\leftrightarrow& G
  \nonumber\\
  \Phi &\leftrightarrow& -\Phi
  \,.
  \nonumber
\end{eqnarray}
This is reminiscent of S-duality, which is known \cite{Duff:1995} to be T-duality in the dual string
picture. We observe that the 2-form background $C$ might have to be identitfied
with the RR 2-form $C_2$ coupled to the D-string (instead of the F-string).

The approach presented here
allows an interesting perspective on superconformal field theories. 
In particular it puts the method of canonical deformations of (S)CFTs in a broader
context and shows how these have to be generalized beyond leading order. 
The concise algebraic form in which it expresses the functional form
of the constraints of such theories suggests itself for further applications.
For instance, as has been discussed in \cite{Schreiber:2003a}, it 
allows the construction of covariant target space Hamiltonians for 
arbitrary backgrounds, a tool that may be helpful for the study of
superstrings in non exactly solvable background fields. In particular,
it would be interesting to study how ghosts and spin fields can be 
incorporated in the present framework and if this would allow to
apply the method of \cite{Schreiber:2003a} to $\mathrm{AdS}_5$ backgrounds. 
Indications how this should work have been given in \cite{Giannakis:2002}.

Finally it may help understand
the spectral approach to SCFTs, as discussed in \S\fullref{deformed inner products on loop space}.

$\,$\\

\acknowledgments
I am grateful to Robert Graham for helpful discussions 
and valuable assistance, and to Ioannis Giannakis for 
informations, discussion and stimulating remarks.
 Furthermore I would like to thank
Eric Forgy for many inspiring exchanges of ideas and thank
Arvind Rajaraman and Lubos Motl for helpful comments. Finally
many thanks to Jacques Distler for setting up the
String Coffee Table.

This work has been supported by the SFB/TR 12.

\newpage
\appendix

\setcounter{equation}{0}
\renewcommand{\theequation}{\Alph{section}.\arabic{equation}}

\section{Canonical analysis of bosonic D1 brane action}
\label{Canonical analysis of bosonic D1 brane action}

The bosonic part of the worldsheet action of the D-string is
\begin{eqnarray}
  {\cal S}
  &=&
    -
    T
    \int
    d^2 \sigma\;
    e^{-\Phi}\sqrt{-{\rm det} (G_{ab} + B_{ab} + \frac{1}{T}F_{ab})}
    +
    T
    \int
    \left(
      C_2 + C_0(B +\frac{1}{T}F)
   \right)
  \,,
\end{eqnarray}
where $G$, $B$, $C_0$ and $C_2$ are the respective background fields and
$F_{ab} = (dA)_{ab}$ is the gauge field on the worldsheet. Indices $a,b$ range
over the worldsheet dimensions and indices $\mu,\nu$ over target space dimensions.

Using \emph{Nambu-Brackets}
$\{X^\mu,X^\nu\} \defas \epsilon^{ab}\partial_a X^\mu \,\partial_b X^\nu$
(with $\epsilon^{01} = 1$, $\epsilon^{ab} = -\epsilon^{ba}$) the term in the
square root can be rewritten as
\begin{eqnarray}
  -{\rm det}\of{G_{ab} + B_{ab} + \frac{1}{T}F_{ab}}
  &=&
  -
  \frac{1}{2}
  \lbrace X^{\mu},X^{\nu}\rbrace
  G_{\mu\mu^\prime}
  G_{\nu\nu^\prime}
  \lbrace X^{\mu^\prime},X^{\nu^\prime}\rbrace
   - (B_{01} + \frac{1}{T}F_{01})^2
  \,.
\end{eqnarray}
The canonical momenta associated with the embedding coordinates $X^\mu$ are
\begin{eqnarray}
  P_\mu &=&
  \frac{\delta {\cal L}}{\delta \dot X^\mu}
  \nonumber\\
  &=&
  T
  \left(
  \frac{1}{e^{\Phi}\sqrt{-{\rm det}(G+B+F/T)}}
  \left(
  X^{\prime \nu}
  G_{\mu\mu^\prime}
  G_{\nu\nu^\prime}
  \lbrace X^{\mu^\prime},X^{\nu^\prime}\rbrace
  +
  B_{\mu\nu}X^{\prime\nu}(B_{01}+ \frac{1}{T}F_{01})
  \right)
  \right)+
  \nonumber\\
  &&
  \;+
  T
  (C_2 + C_0 B){}_{\mu\nu}X^{\prime\nu}
  \,.
  \nonumber\\
\end{eqnarray}
On the other hand the canonical momenta associated with the gauge field read
\begin{eqnarray}
  \label{D1 gauge field canonical momenta}
  E_0
  &\defas&
  \frac{\delta {\cal L}}{\delta \dot A_1}
  \;=\;
  0
  \nonumber\\
  E_1
  &\defas&
  \frac{\delta {\cal L}}{\delta \dot A_1}
  \;=\;
  \frac{1}{e^{\Phi}\sqrt{-{\rm det}(G+B+F/T)}}
  (B_{01}+ \frac{1}{T}F_{01})
  +
  C_0
  \,.
\end{eqnarray}
Since the gauge group is ${\rm U}\of{1}$, $A_\mu$ is a periodic variable and hence the 
eigenvalues of $E_1$ are discrete \cite{Witten:1995}:
\begin{eqnarray}
  E_1 &\defas& p \,\in \Z
  \,.
\end{eqnarray}
Inverting \refer{D1 gauge field canonical momenta} 
allows to rewrite the canonical momenta $P_\mu$ as
\begin{eqnarray}
  P_\mu &=&
  \frac{1}{\sqrt{-\mathrm{det}\of{G}}}
  \tilde T
  X^{\prime \nu}
  G_{\mu\mu^\prime}
  G_{\nu\nu^\prime}
  \lbrace X^{\mu^\prime},X^{\nu^\prime}\rbrace
  +
  T
  (C_2 + p B){}_{\mu\nu}X^{\prime\nu}
  \,,
\end{eqnarray}
where
\begin{eqnarray}
  \tilde T
  &\defas&
  T\sqrt{e^{-2\Phi} + (p-C_0)^2}
\end{eqnarray}
is the tension of a bound state of one D-string with p F-strings \cite{IgarishiItohMamimuraKuriki:1998}.
In this form it is easy to check that the following two constraints are satisfied:
\begin{eqnarray}
  \label{D-string constraints}
  &&(P-T(C_2 + p B)\inner X^\prime) \inner
  (P-T(C_2 + p B)\inner X^\prime)
  +
  \tilde T^2
  X^\prime \inner X^\prime
  = 0 
  \nonumber\\
  &&(P-T(C_2 + p B)\inner X^\prime)\inner X^\prime = 0
  \,,
\end{eqnarray}
which express temporal and spatial reparameterization invariance, respectively.
For constant $\tilde T$ this differes from the familiar constraints for the pure F-string 
only in a redefinition of the tension and the couplings to the background 2-forms.

For non-constant $\tilde T$, however, things are a little different. 
For the purpose of comparison with the results in \S\fullref{Dilaton background}
consider the case $B = C_0 = C_2 = p = 0$ and $\Phi$ possibly non-constant. In this case the constraints
\refer{D-string constraints} can be equivalently rewritten as
\begin{eqnarray}
  \mathcal{P_\pm}^2 &=& 0
\end{eqnarray}
with
\begin{eqnarray}
  \label{Born-Infeld current for pure dilaton background}
  {\cal P}_{\mu,\pm}
  &=&
  e^{\Phi/2}
  P_\mu 
  \pm
  T e^{-\Phi/2}
   G_{\mu\nu} 
   X^{\prime\nu}
  \,.
\end{eqnarray}
Up to fermionic terms this is the form found in \refer{Phi Dirac operators}.

\newpage
\bibliography{std}

\end{document}